\pdfoutput=1
\documentclass[aip,cha,amsmath,amssymb,reprint,author-numerical]{revtex4-1}

\usepackage{graphicx}
\usepackage{dcolumn}
\usepackage{bm}
\usepackage{hyperref}
\newcommand{\ohm}{$\mathrm{\Omega}$}
\newcommand{\ep}{$\epsilon$}

\begin{document}

\preprint{AIP/123-QED}

\title{Amplitude death and synchronized states in nonlinear time-delay systems coupled through mean-field diffusion} 



\author{Tanmoy Banerjee}
\email{tbanerjee@phy.buruniv.ac.in}
 \email{tanbanrs@yahoo.co.in}
\affiliation{Department of Physics, University of Burdwan, Burdwan 713 104, West Bengal, India.}
\author{Debabrata Biswas}
\affiliation{Department of Physics, University of Burdwan, Burdwan 713 104, West Bengal, India.}


\received{:to be included by reviewer}
\date{\today}

\begin{abstract}
We explore and experimentally demonstrate the phenomena of amplitude death (AD) and the corresponding transitions through synchronized states that lead to AD in coupled {\it intrinsic time-delayed} hyperchaotic oscillators interacting through mean-field diffusion. We identify a novel synchronization transition scenario leading to AD, namely transitions among AD, generalized anticipatory synchronization (GAS), complete synchronization (CS), and generalized lag synchronization (GLS). This transition is mediated by variation of the difference of {\it intrinsic time-delays} associated with the individual systems, and has no analogue in non-delayed systems or coupled oscillators with {\it coupling time-delay}. We further show that, for equal intrinsic time-delays, increasing coupling strength results in a transition from the unsynchronized state to AD state via in-phase (complete) synchronized states. Using Krasovskii--Lyapunov theory, we derive the stability conditions that predict the parametric region of occurrence of GAS, GLS, and CS; also, using a linear stability analysis we derive the condition of occurrence of AD. We use the error function of proper synchronization manifold and a modified form of the similarity function to provide the quantitative support to GLS and GAS. We demonstrate all the scenarios in an electronic circuit experiment; the experimental time-series, phase-plane plots, and generalized autocorrelation function computed from the experimental time series data are used to confirm the occurrence of all the phenomena in the coupled oscillators.
\end{abstract}


\maketitle 

\begin{quotation}
Coupled dynamical systems show a plethora of complex collective behaviors like synchronization, phase-locking, amplitude death, etc. Amplitude death (AD) is one of the intriguing phenomena that occurs in coupled oscillators when they  interact in such a way as to suppress each other’s oscillations and collectively go to a stable fixed point. To induce AD, Mean-field diffusive coupling is an important coupling scheme, because it removes the constraint of having parameter mismatch or time-delay coupling to obtain AD. Further, several transitions leading to AD have been identified and investigated. Surprisingly, very few works are reported on AD in systems with {\em intrinsic time-delay}, and most of them consider a delayed coupling scheme. Dynamical systems having intrinsic time-delay are infinite dimensional and very complex, and thus they have to be treated separately. Further, practical implementation of intrinsic time-delay systems are difficult and challenging, thus studies on AD in these systems and its experimental demonstration are of considerable importance. In this paper, we extensively study the dynamical behavior of {\em intrinsic time-delayed} hyperchaotic oscillators coupled with mean-field diffusion; we theoretically explore and experimentally demonstrate the phenomena of AD, and a  new transition scenario that leads to AD, namely the transitions among AD, generalized (anticipatory, lag) synchronization, and complete synchronization.
\end{quotation}

\section{Introduction}
\label{sec:intro}
Cooperative phenomena in coupled dynamical systems are of significant interest in the field of physical science, biological science, and engineering applications \cite{piko}. The prominent cooperative behaviors that occur in periodic and chaotic oscillating systems are synchronization \cite{piko}, amplitude death\cite{prasad1,*prasad1b}, phase-flip transition\cite{dana1,*dana0}, etc. {\it Amplitude death} (AD) is one of the fascinating and important emergent phenomena in which quenching of amplitude or cessation of oscillation to the steady state occurs in coupled dynamical systems under some proper parametric conditions. Studies on AD has been attracting the attention of the researchers for more than two decades owing to its importance in the field of physical science, biology, oceanography, etc\cite{st2,prasad1,*prasad1b}. The phenomenon of AD has first been reported by \citet{yamaAD}; later the same has been studied in detail by \citet{bareli}, and \citet{shiino}. Recently, an extensive review on AD has been reported in Ref.\onlinecite{prasad1,*prasad1b} that discussed several aspects of AD, and a thorough literature review. 

To induce AD in coupled oscillators, several coupling schemes have been proposed (see Ref.\onlinecite{prasad1,*prasad1b} and references therein); a few of them are linear diffusive coupling\cite{prasad1}, dynamic coupling \cite{prasad1}, environmental coupling \cite{resmi12,*sharma2}, etc. In most of the coupling schemes, to induce AD, it is necessary that the systems are mismatched. In a seminal paper by \citet{asen}, for the first time, it has been shown that a {\it coupling time-delay} can induce AD even in identical limit cycle oscillators (i.e., oscillators without parameter mismatch). This paper leads to  a whole lot of research activity in inducing AD, e.g., in electronic circuit\cite{asen1}, in an assembly of delay coupled oscillator\cite{asen2}, distributed delay oupled oscillator \cite{atay}, ring of delay coupled limit cycle oscillator\cite{asen3}, delay coupled chaotic oscillator\cite{prasad3}, and delay coupled nonidentical oscillators\cite{yao1}, to name a few. Later, the constraints of having either parameter missmatch or coupling time-delay to induce AD have been removed with dynamic coupling, conjugate coupling \cite{karna}, and linear augmentation \cite{sharma} coupling, where AD occurs in identical systems with instantaneous coupling. Recently, another interesting process of inducing AD in {\it identical} systems has been reported by \citet{Shrimali12} that considers {\it mean-field} (MF) diffusive coupling. Earlier, AD through MF coupling has been reported in Ref. \onlinecite{asen2} and Ref.\onlinecite{st,*de} with  coupling time-delay, and distributed frequencies or parameter mismatches, respectively.

Another important topic of study in the context of AD is the {\it transition scenarios leading to AD}. In delay coupled periodic and chaotic oscillators (without intrinsic time-delay), several transition scenarios have been reported; The most prominent are, (a) phase-flip transition\cite{prasad1}, and  (b) transitions from chaotic to AD state via quasiperiodic and periodic states\cite{prasad3}, \cite{choi}. It has been shown in Ref.\onlinecite{phflip} that phase-flip transition, i.e., the abrupt change from in-phase synchronized dynamics to antiphase synchronized dynamics, is caused by the time-delay in the {\it coupling path}; the same is  true for the latter transition, also. In the non-delayed coupling case, {\it coupling strength} is the defining parameter; depending upon coupling strength, transition from unsynchronized states to AD via in-phase (anti-phase) synchronized states has been reported in Ref.\onlinecite{resmi12,*sharma2} (environmental coupling) and Ref.\onlinecite{Shrimali12} (mean-field couling).  But, the effect of {\it intrinsic time-delay} on the transition scenario is yet to be explored.

Surprisingly, in all the above mentioned works on AD, the coupled dynamical systems are considered to be periodic or chaotic oscillators with no {\it intrinsic time-delay}; in these studies, when present, time-delay appears only in the coupling path. The systems with intrinsic  time-delay are infinite dimensional and very complex\cite{ataybook,*lakbook}. Recent ongoing interest in intrinsic time-delay systems originates from the fact that  in real world often we have to encounter with the systems having {\it intrinsic time-delay}; examples include, blood production in patients with leukemia (Mackey-Glass model) \cite{mac}, dynamics of optical systems (e.g. Ikeda system) \cite{ikeda1}, population dynamics \cite{pop}, El Ni\~{n}o/southern oscillation (ENSO) \cite{nino}, etc. Due to infinite-dimensionality of the time-delay systems, studies on their collective behaviors like synchronization and AD are much more involved, and need special attention \cite{dana}. In this context, synchronization of chaos and hyperchaos in intrinsic time-delay systems is a well explored topic and extensive research has already been devoted\cite{py,*sahalag,*sahaas,*saha,*tangs,*lakps,*lakpsexpt,*banerjee13} but the same is not true for AD.

Contrary to  the phenomenon of AD in low dimensional systems with {\it coupling time-delay}, amplitude death in systems with {\it intrinsic time-delay} is a less explored topic. The first observation of AD in {\it intrinsic time-delayed oscillators} (i.e., oscillator with {\it intrinsic time-delay}) has been reported by \citet{kon1} in which dynamic and delayed couplings were studied. AD in networks of delay-coupled delay oscillators has been studied analytically in Ref.\onlinecite{hefner}. Ref.\onlinecite{kon2} reports AD in intrinsic time-delayed oscillators coupled via multiple delay connections. Recently, an important technique of inducing AD has been proposed in Ref.~\onlinecite{dana}, which employ time-delayed open-plus-closed-loop coupling. In all of these works (except the case of dynamic coupling), AD is mediated by the presence of moderate or very long time-delay in the coupling path. The presence of time-delay in coupling path makes the dynamics of the coupled systems more complex, and at the same time much difficult for analysis and practical implementation.  

In this paper, our aim is to study amplitude death and the corresponding synchronization transitions leading to AD  in coupled {\it intrinsic time-delayed}  hyperchaotic oscillators interacting through mean-field diffusive coupling. That is, the time-delay we are dealing with is the {\it intrinsic time-delay} associated with the individual systems, not the {\it coupling time-delay}.  We identify a {\it novel} synchronization transition scenario that leads to AD, namely the transitions among AD,  {\it generalized anticipatory synchronization} (GAS), complete synchronization (CS), and {\it generalized lag synchronization} (GLS); this transition occurs for the variation of the difference of {\it intrinsic time-delays}, $\tau_d=(\tau_2-\tau_1)$ ($\tau_1$ and $\tau_2$ are the intrinsic time-delays associated with two coupled systems).  We show the occurrence of GAS for $\tau_d<0$, CS for $\tau_d=0$, and GLS for $\tau_d>0$. To the best of our knowledge, this type of transition has not been reported in the literature yet;  also, it has no counterpart in oscillators with no intrinsic time-delay (with or without couping time-delay). Contrary to the transitions in oscillators with coupling time-delay (see Senthilkumar and Lakshmanan of Ref. 28 that deals with the interplay of intrinsic
and coupling delay), the transition reported here is related to the {\it intrinsic time-delay} associated with the individual systems.  By definition\cite{gls1,*gls2}, in the GAS, CS, and GLS conditions, there exists a smooth function $H$ such that $x_2(t)=H(x_1(t+\tau))$, $\tau\in \mathbb{R}$; for GAS, $\tau>0$, and for GLS, $\tau<0$; for conventional AS and LS states $H$ is an identity function, i.e., $x_2(t)=x_1(t+\tau)$, and as usual, for a CS state, $x_2(t)=x_1(t)$. Earlier, it has been shown that  mismatch in intrinsic time-delay in linearly coupled time-delay systems gives rise to generalized synchronization (GS) \cite{saha,tangs}. In non-delayed systems, to observe GLS and GAS, appropriate controller has to be designed; Ref.\onlinecite{gls1,*gls2} reported two such controller design techniques to induce GLS and GAS in low dimensional systems under {\it drive-response} coupling. Experimental confirmation of GLS has been reported in Ref.\onlinecite{glsexpt} which consider R\"{o}ssler systems. Unlike non-delayed systems, in our present case no controller is required but only variations of the intrinsic time-delay give rise to GAS and GLS. Further, at present, there exists no general theory or confirmatory quantitative measures of GAS and GLS. In this paper we derive a general stability analysis (under some constraints) for the GAS and GLS states using Krasovskii--Lyapunov theory. Also, we use error function and a modified form of the similarity function to provide the quantitative support to the occurrence of GAS and GLS. 

We further study the effect of coupling  parameter for equal intrinsic time-delays. It is shown that depending upon the coupling strength and mean-field parameters, the coupled systems show a transition from the unsynchronized state to AD state via in-phase and complete synchronized states.  The occurrence of CS and AD is predicted analytically using Krasovskii--Lyapunov theory and linear stability analysis, respectively.

To exemplify our study, we consider a prototype hyperchaotic system with intrinsic  time-delay, recently proposed in Ref. \onlinecite{banerjee12}. We compute the Lyapunov exponent (LE) spectrum of the coupled systems, correlation of probability of recurrence, cross correlation functions, and eigenvalue spectrum to identify the synchronized states and AD state in the parameter space.  Finally, we set an electronic circuit experiment to demonstrate all the transition scenarios. Experimental waveforms and phase-plane plots are used to visualize the transitions.  We show that analytical and numerical results agree well with the experimental observations. 

The rest of this paper is organized in the following way: The next section describes the mathematical model of the time-delay systems under mean-field coupling. Stability analysis is reported in Section \ref{sec:stab}. Numerical simulation results are described in Section \ref{sec:num}. Section \ref{sec:expt} reports the experimental implementation of the coupled system and experimental demonstration of amplitude death. Finally, we summarize the main observations of our study in Section \ref{con}.
\section{Mean-field coupling}
We consider $N$ number of first-order time-delay dynamical systems interacting through mean-field diffusive coupling; mathematical model of the coupled system is given by
\begin{equation}\label{x1x2} 
\dot{x_i}=h(x_i,x_{i\tau_i};p)+\epsilon\left(Q\overline{X}-x_i\right),
\end{equation}
with $i=1\cdots N$, $\overline{X}=\frac{1}{N}\sum_{i=1}^{N}x_i$ is the mean-field of the coupled system. $x_{\tau}\equiv x(t-\tau)$, $\tau\in \mathbb{R}^+$ is the constant time-delay, and $p$ represents the $m$ dimensional parameter space. The coupling strength is given by $\epsilon$, and $Q$ is a control parameter that determines the density of mean-field \cite{Q,Shrimali12} ($0\le Q\le1$). Here the function $h(x_i,x_{i{\tau_i}};p)$ is given by $h(x_i,x_{2{\tau_i}},p)=-ax_i-b_if(x_i{_{{\tau}_i}})$, thus the individual units are represented by the following scalar first-order, retarded type delay differential equations:
\begin{equation}\label{sys}
\dot{x_i}=-ax_i-b_if(x_i{_{{\tau}_i}}),
\end{equation}
where $a>0$ and $b_i$ are the system parameters, and  $\tau_i$ is the intrinsic time-delay associated with the individual systems. Eq.\eqref{sys} represents a general class of first-order, nonlinear, retarded delay-differential equations. For example, for Mackey-Glass system \cite{mac}: $f(x_i{_{{\tau}_i}})=-\frac{x_{\tau_i}}{1+x_{\tau_i}^c}$; for Ikeda system \cite{ikeda1}: $f(x_i{_{{\tau}_i}})=\sin(x_{i\tau_i})$, etc. Thus, Eq.\eqref{x1x2} represents mean-field diffusive coupling scheme for any first-order delay dynamical systems.

\section{\label{sec:stab}Stability analysis}
In this section we analyze the asymptotic stability of the synchronization of the coupled systems given in Eq. \eqref{x1x2}. Here we restrict our study to a pair ($N=2$) of time-delay systems.
\subsection{\label{sub:cs}Krasovskii--Lyapunov theory: complete synchronization ($\tau_1=\tau_2$)}
Let us define the error function as $\Delta=(x_1-x_2)$, and also let $\tau_1=\tau_2=\tau$. Time evolution of the error function that describes the error dynamics of \eqref{x1x2} is given by
\begin{equation}
\dot{\Delta}=-(a+\epsilon)\Delta-(b_1-b_2)f(x_1{_{\tau}})-b_2f'(x_1{_{\tau}})\Delta_{\tau}.\label{ddt}
\end{equation}
This is an inhomogeneous equation and difficult to deal with; to make it homogeneous we impose the following constraint: $b_1=b_2=b$, which is the necessary condition of complete synchronization. Now, Eq. \eqref{ddt} becomes
\begin{equation}
\dot{\Delta}=-(a+\epsilon)\Delta-bf'(x_1{_{\tau}})\Delta_{\tau}.\label{deldt}
\end{equation}
According to the Krasovskii--Lyapunov theory \cite{Kra03}, a stable synchronization implies the stability of the origin of \eqref{deldt}. The sufficient condition for the stability of synchronization requires the definition of a positive definite functional, $V(t)$, given by
\begin{equation}
V(t)=\frac{1}{2}\Delta^2+\mu\int_{-\tau}^0{\Delta^2(t+\varphi)d\varphi}.\label{v}
\end{equation} 
Here $\mu>0$ is an arbitrary positive parameter. The stability of the origin of \eqref{deldt} requires that the time derivative of $V(t)$ be negative. Now,
\begin{equation}
\frac{dV}{dt}=-\mu\Delta^2\Gamma(X,\mu),\label{dv}
\end{equation}
where $\Gamma(X,\mu)=\frac{(a+\epsilon-\mu)}{\mu}+\frac{bf'(x_{1_{\tau}})}{\mu}X+X^2$ and $X=\frac{\Delta_\tau}{\Delta}$. Thus, from Eq.\eqref{dv} it may be noted that the negativity of $dV/dt$ requires the following condition to be valid: $\Gamma_{min}>0$. Now, $\Gamma_{min}$ is derived as
\begin{equation}
\Gamma_{min}=\frac{4\mu(a+\epsilon-\mu)-b^2f'^2(x_{1_\tau})}{4\mu^2}.\label{g}
\end{equation}
Hence $\Gamma_{min}>0$ implies that
\begin{equation}
a+\epsilon>\frac{b^2f'^2(x_{1_\tau})}{4\mu}+\mu=\Phi(\mu).\label{phi}
\end{equation}
Here $\Phi(\mu)$ is a function of $\mu$. Now, we find the minimum value $\Phi_{min}$ by setting $\frac{d\Phi}{d\mu}=0$. That gives $\mu=\frac{\lvert bf'(x_{1_\tau})\rvert}{2}$. With this value of $\mu$, one gets the minimum value of $\Phi$ as $\Phi_{min}=\lvert bf'(x_{1_\tau})\rvert$; using this in Eq.\eqref{phi} we get the following sufficient condition of complete synchronization:
\begin{equation}
a+\epsilon>\lvert bf'(x_{1_\tau})\rvert.\label{cond}
\end{equation}
Note that, Eq.\eqref{cond} represents the sufficient condition of complete synchronization for {\it any} general first-order time-delay systems of the form given by Eq.\eqref{sys} coupled via mean-field diffusion. 
\subsection{Generalized (anticipatory, lag) synchronization: $\tau_1\ne\tau_2$}
For the GAS, GLS cases we consider the following error function: $\Delta=H(x_{1_{\tau_2-\tau_1}})-x_2$, where $x_{1_{\tau_2-\tau_1}}=x_1(t-(\tau_2-\tau_1))$. Using this we can express three different synchronization phenomena, namely, generalized (anticipatory, lag), and complete synchronization. GAS is observed for $\tau_1>\tau_2$; under this condition one has $x_2(t)=H(x_1(t+\lvert\tau_2-\tau_1\rvert))$. For $\tau_1=\tau_2$ we have CS, i.e., $x_2(t)=x_1(t)$. GLS occurs for $\tau_1<\tau_2$; in this case one has $x_2(t)=H(x_1(t-\lvert\tau_2-\tau_1\rvert))$.

The time evolution of the error function is given by: $\dot{\Delta}=\dot{H}(x_{1_{\tau_2-\tau_1}})-\dot{x_2}$. Since $H$ is an unknown, arbitrary function, further analysis is not possible. Considerable progress can be made if we consider $H(u)=\Psi u$ where $\Psi$ is an appropriate scaling factor; this is a valid approximation only in the strong coupling case where the dynamics becomes periodic. With this we have
\begin{equation}\label{deldot}
\begin{split}
\dot{\Delta}=-(a+\epsilon(1-\frac{Q}{2}))\Delta+b(f(x_2(t-\tau_2)-f(x_2(t-\tau_1))\\
-bf'(x_2(t-\tau_1))\Delta_{\tau_{1}}+\frac{\epsilon Q}{2}(\Psi x_2(t-(\tau_2-\tau_1))-x_1),
\end{split}
\end{equation}
where, $\Delta_{\tau_1}=\Delta(t-\tau_1)$, and $b_1=b_2=b$ . The synchronization manifold is locally attracting if the origin of \eqref{deldot} is stable. It can be noted that for $\tau_1=\tau_2$, i.e. complete synchronization, $\Psi=1$, and $x_2(t-(\tau_2-\tau_1))-x_1=-\Delta$; thus Eq.\eqref{deldot} reduces to Eq.\eqref{deldt}, and the Krasovskii--Lyapunov theory gives the same result as Eq.\eqref{cond}.  For $\tau_1\ne\tau_2$, since Eq.\eqref{deldot} is an inhomogeneous equation it is not tractable for further analysis; but, in the small intrinsic time-delay difference condition (i.e. $\lvert\tau_2-\tau_1\rvert$ is small), we can neglect the second term in Eq.\eqref{deldot}, and also write $\left(\Psi x_2(t-(\tau_2-\tau_1))-x_1\right)=-\Psi'\Delta$, where $\Psi'$ is a new scaling factor (note that for $\lvert\tau_2-\tau_1\rvert=0$, $\Psi'=1$); under this condition Eq.\ref{deldot} reduces to 
\begin{equation}\label{deldot1}
\dot{\Delta}=-\left(a+\epsilon(1-\frac{Q}{2}(1-\Psi')\right)\Delta -bf'(x_2(t-\tau_1))\Delta_{\tau_{1}}.
\end{equation}
Note that, Eq.\eqref{deldot1} has the same form as Eq.\eqref{deldt}; using the Krasovskii--Lyapunov theory and the same arguments of the previous subsections we arrive at the following stability condition for the generalized (anticipatory, lag) synchronization:
\begin{equation}\label{cond1}
a+\epsilon(1-\frac{Q}{2}(1-\Psi'))>\lvert bf'(x_2(t-\tau_1))\rvert.
\end{equation}
\subsection{\label{sub:ad} Linear stability analysis: amplitude death}
Next, to find out the condition of amplitude death we analyze the stability of synchronization by considering the deviations from the synchronized state. The same for the low-dimensional systems (without intrinsic time-delay) has been reported in Ref.\onlinecite{resmi12,*sharma2} and Ref.\onlinecite{Shrimali12}. Let us define $\psi$ and $\phi$ to be the deviations from the synchronized states of the system variables $x_1$ and $x_2$ in \eqref{x1x2}, respectively. Then the linearization of the system along these deviations gives
\begin{subequations}\label{lineq}
\begin{eqnarray}
\dot{\psi}&=-a\psi-b_1f^{\prime}(x_{1_\tau})\psi_\tau+\epsilon\big(Q\frac{\psi+\phi}{2}-\psi\big),\label{lineq1}\\
\dot{\phi}&=-a\phi-b_2f^{\prime}(x_{2_\tau})\phi_\tau+\epsilon\big(Q\frac{\psi+\phi}{2}-\phi\big).\label{lineq2}
\end{eqnarray} 
\end{subequations}   
An exact analysis of Eq.\eqref{lineq} is not possible due to the presence of the delay term, which makes the characteristic equation a quasi-polynomial one. We consider $b_1=b_2$, and for the the complete synchronization we have $x_1=x_2$ and $x_{1\tau}=x_{2\tau}=x_{\tau}$. Let us define $g(x_{\tau},\psi,\psi_\tau)\equiv-a\psi-bf^{\prime}(x_\tau)\psi_\tau$, and $g(x_{\tau},\phi,\phi_\tau)\equiv-a\phi-bf^{\prime}(x_\tau)\phi_\tau$. Now, Eq.\eqref{lineq} reduces to
\begin{subequations}\label{lineqg}
\begin{eqnarray}
\dot{\psi}&=g(x_\tau,\psi,\psi_\tau)+\epsilon\big(Q\frac{\psi+\phi}{2}-\psi\big),\label{lineqg1}\\
\dot{\phi}&=g(x_\tau,\phi,\phi_\tau)+\epsilon\big(Q\frac{\psi+\phi}{2}-\phi\big).\label{lineqg2}
\end{eqnarray}
\end{subequations} 
The Jacobian matrix of the system is described by
\begin{equation}\label{cheq}
\left(
\begin{array}{cc}
\delta+\epsilon\big(\frac{Q}{2}-1\big) & \frac{Q\epsilon}{2}\\
\frac{Q\epsilon}{2} & \delta+\epsilon\big(\frac{Q}{2}-1\big)
\end{array}
\right)=0,
\end{equation}
where, we consider that the time-averaged values of $g^{\prime}(x_\tau,\psi,\psi_\tau)$ and $g^{\prime}(x_\tau,\phi,\phi_\tau)$ are approximately same and are equal to an effective constant $\delta$.  This type of approximation has been used in Refs.\onlinecite{Shrimali12}, \onlinecite{Resmi10,*Resmi11}  for the low-dimensional systems without intrinsic time-delay; here we extend the same for the time-delay systems. 

Now the characteristic equation of the Jacobian matrix \eqref{cheq} is
\begin{equation}\label{ch}
\lambda^2-2\Lambda\lambda+\Lambda^2-\bigg[\frac{\epsilon Q}{2}\bigg]^2=0,
\end{equation}
where, $\Lambda=\delta+\epsilon\big(\frac{Q}{2}-1\big)$. Thus, we have the following two eigenvalues: $\lambda_1=\delta+\epsilon(Q-1),~~ \lambda_2=\delta-\epsilon$. For amplitude death to occur, $\lambda_{1,2}$ should be negative\cite{Shrimali12}, which gives: $Q<1-\frac{\delta}{\epsilon}$, and $\epsilon>\delta$. Thus the critical parametric condition for which amplitude death occurs is given by
\begin{equation}\label{ad}
Q_{cr}=1-\frac{\delta}{\epsilon_{cr}},
\end{equation} 
along with $\epsilon_{cr}>\delta$; here $Q_{cr}$ and $\epsilon_{cr}$ are the critical values of the mean-field parameter and coupling strength, respectively. 

\section{\label{sec:num}Numerical simulation}
\subsection{System description}
\label{sec:sysdes}
For numerical verification of the analytical predictions and demonstration of the collective behaviors, we consider the following first-order nonlinear retarded time-delayed system recently proposed in Ref.\onlinecite{banerjee12}:
\begin{equation}
\dot{x}=-ax-bf(x_{\tau}),\label{syseq}
\end{equation}
where $a$ and $b$ are positive parameters. The nonlinear function $f(x_{\tau})$ is given by
\begin{equation}
f(x_{\tau})=-0.5n(\lvert x_{\tau}\rvert+x_{\tau})+m\tanh{(lx_{\tau})},\label{hw}
\end{equation}
where, $n$, $m$, and $l$ are positive parameters that determine the nature of the nonlinearity. There exists a large number of choices of these parameters for which chaos and hyperchaos can be observed \cite{banerjee12}.

\begin{figure}
\includegraphics[width=.48\textwidth]{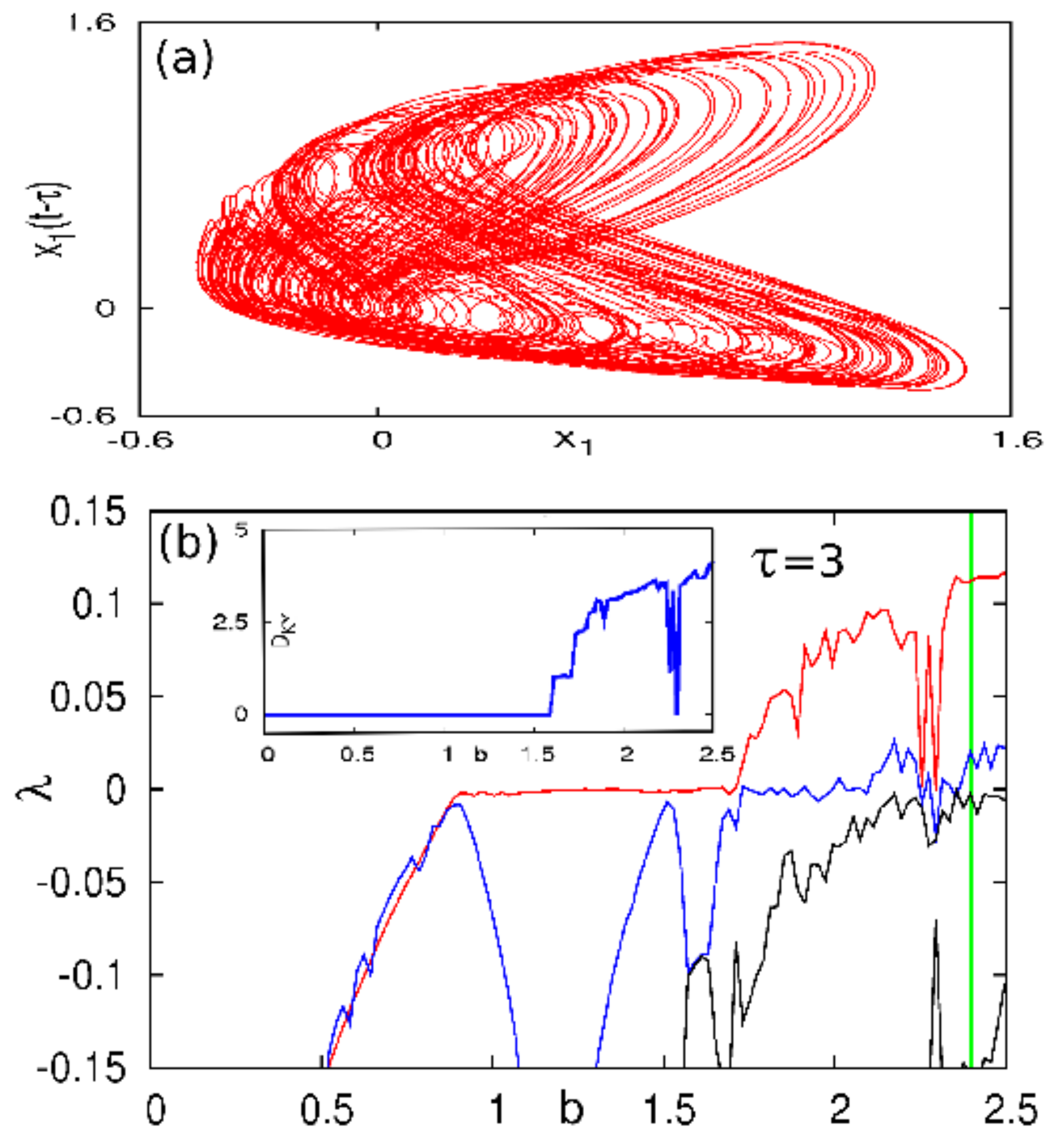}
\caption{\label{phlpv} (a) Hyperchaotic attractor for $b=2.4$ and $\tau=3$. (b)~Lyapunov exponent spectrum with $b$ for $\tau=3$, the vertical line is for $b=2.4$; the inset shows Kaplan-York dimension ($D_{KY}$). Other parameter values are: $a=1$, $n=1.15$, $m=0.97$, $l=2.19$.}
\end{figure}

The detailed chaotic and hyperchaotic dynamical behaviors have been reported in Ref.~\onlinecite{banerjee12}. The system \eqref{syseq} (with \eqref{hw}) has only one (trivial) fixed point at $x^{*}=0$. It has been shown that, keeping $b$ fixed, if one varies $\tau$, the system shows a period doubling route to chaos and hyperchaos (parameters are: $a=1$, $n=1.15$, $m=0.97$, $l=2.19$). For example, for $b=2.4$, at $\tau\approx0.59$, the fixed point loses its stability through Hopf bifurcation and a stable limit cycle appears; after a period doubling sequence, chaos and hyperchaos are observed at $\tau \approx 1.61$ and $\tau \approx2.54$, respectively. The equivalent period-doubling route has been observed if one keeps $\tau$ fixed and varies $b$; e.g., for $\tau=3$, at $b\approx0.92$ oscillation set in; chaos occurs for $b\approx1.71$ and hyperchaos occurs for $b\approx2$. Figure \ref{phlpv}(a) shows the hyperchaotic attractor for $b=2.4$ and $\tau=3$. Figure \ref{phlpv}(b) shows the Lyapunov exponent spectrum (along with the Kaplan-York dimension) in the $b$ parameter space for $\tau=3$; the presence of multiple positive Lyapunov exponents along with the strange attractor ensures the occurrence of hyperchaos in the system.
\subsection{Numerical results}
The system equation \eqref{x1x2} (with \eqref{syseq} and \eqref{hw}) is simulated numerically using Runge--Kutta algorithm with step size $h=0.01$. The following initial functions have been used for all the numerical simulations: for the $x_1$-system: $\xi_{x1}(t)=0.95$, and for the $x_2$-system: $\xi_{x2}(t)=0.85$. Also, the following system design parameters are chosen throughout the numerical simulations: $a=1$, $n=1.19$, $m=0.97$, $l=2.19$, and $b_1$=$b_2$=$2.4$.

\subsubsection{Effect of intrinsic time-delay: transitions among AD, GAS, CS, and GLS.}
\begin{figure*}
\includegraphics[width=.7\textwidth]{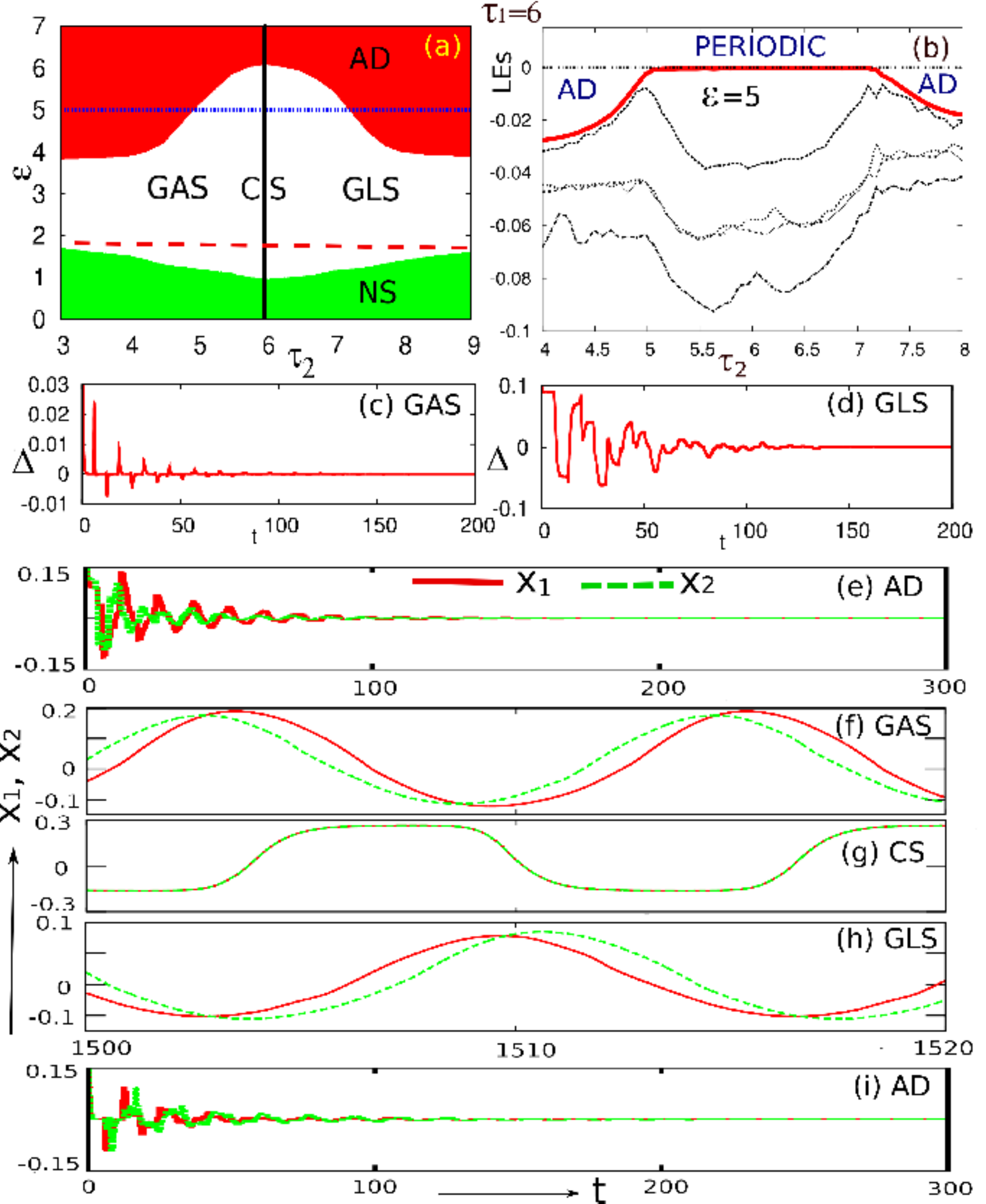}
\caption{\label{epstau} (Color online) $\mathbf{\tau_1}=6$. {\bf (a)} Phase diagram in {\ep}--$\tau_2$ parameter space ($Q=0.6$). NS: unsynchronized state. Horizontal dashed (red) line indicates the analytically obtained critical curve for obtaining synchronized states. {\bf (b)} The Lyapunov exponent (LE) spectrum with $\tau_2$ for $\epsilon=5$, $\tau_1=6$, and $Q=0.6$. {\bf (c, d)} Time evolution of the error function of GAS ($\tau_2$=5.8) (c), and GLS ($\tau_2=6.2$) (d); both show that synchronization error asymptotically goes to zero. {\bf (e--i)} time series of $x_1(t)$ and $x_2(t)$ show transitions from AD ($\tau_2=4$) to GAS ($\tau_2=5.2$) to CS ($\tau_2=6$) to GLS ($\tau_2=7$), and finally again to AD ($\tau_2=8$).}
\end{figure*}

At first, we explore the effect of {\it intrinsic time-delay} on the dynamics of the coupled system. Figure \ref{epstau}(a) depicts the phase diagram showing the zone of unsynchronized, synchronized, and AD states in $\epsilon-\tau_2$ space for a constant $\tau_1$.  We observe that beyond a certain coupling strength (e.g., $\epsilon=5$, along the horizontal dotted (blue) line of Fig.\ref{epstau}(a)), for a fixed $\tau_1$, if $\tau_2$ is varied from a low to high value the coupled systems show transitions from AD to {\it generalized anticipatory synchronization} (GAS)(for $\tau_1>\tau_2$) to complete synchronization (for $\tau_1=\tau_2$) to {\it generalized lag synchronization} (GLS) (for $\tau_1<\tau_2$), and again to AD state. Further, for a weaker coupling strength  we have a transition from GAS to CS to GLS, and no AD occurs. The analytically obtained critical value of $\epsilon$, $\epsilon_{cr}$, beyond which synchronization occurs is shown in Fig.\ref{epstau}(a) with dashed (red) line, which is obtained by using Eq.\eqref{cond1} (with, $\tau_1=6, \Psi'=1.1, Q=0.6$); it lies well within the numerically obtained synchronized zone indicating the effectiveness of our stability analysis. 

Next, we consider $\epsilon=5$ and vary $\tau_2$ (i.e., along the dotted (blue) line of Fig.\ref{epstau}(a)). Figure \ref{epstau}(b) shows the first five LEs; with increasing $\tau_2$, the largest LE (solid(red) line) makes a transition from negative values (indicating AD) to zero value (indicating periodic and synchronized (since all other LEs are negative) states), and again to negative values (indicating AD state). From the LE spectrum it is also obvious that, sufficient mismatch in intrinsic time-delays enhance the region of AD in the parameter space. Further, unlike {\it delay coupled oscillators}, we find no ``avoided crossing"\cite{phflip} in the LE spectrum confirming that no {\it phase-flip transition} occurs for the variation of intrinsic time-delay. We compute $\Delta$ from Eq.\eqref{deldot1} to show the real time variation of the error function of GAS and GLS. Figure \ref{epstau}(c) and (d) show this for the GAS ($\tau_2=5.8$) and GLS ($\tau_2=6.2$), respectively with $\epsilon=1.8$ which is greater than the analytically obtained value of $\epsilon_{cr}$ (with, $\tau_1=6, \Psi'=1.1, Q=0.6$). It is clear that the  error function attains a zero steady state value confirming the occurrence of GAS and GLS. 

Next, for $\tau_1=6$ and $\epsilon=5$, we plot the time evolution of $x_1(t)$ (solid(red) line) and $x_2(t)$ (dotted(green) line). With the variation of $\tau_2$ we can see the transitions (Fig.~\ref{epstau}(e)--(i)) from AD ($\tau_2=4$) to GAS ($\tau_2=5.2<\tau1$), CS ($\tau_2=6=\tau_1$) to GLS ($\tau_2=7>\tau_1$), and finally again to AD ($\tau_2=8$). In the transient regions of the AD states in Fig.~\ref{epstau} (e) and (i) one can observe that the GAS and GLS behaviors, respectively, lead to AD; we find no ``phase-flip" in the transient behaviors for any intrinsic time-delay; this along with the LE spectrum confirms that variation of intrinsic time-delay does not result in {\it phase-flip transition}.

Since at present, there exists no confirmatory quantitative measure of GAS and GLS, we compute a modified form of the similarity function $S$ defined as\cite{glsexpt}
\begin{equation}\label{sim}
S^{2}(\tau_d)=\frac{\langle[H(x_1(t+\tau_d))-x_2(t)]^2\rangle}{[\langle x_{1}^{2}(t)\rangle\langle x_{2}^{2}(t)\rangle]^\frac{1}{2}},
\end{equation} 
where $\tau_d$ is the time-delay between $x_1$ and $x_2$ that is equal to $\tau_d=\lvert\tau_2-\tau_1\rvert$. For the synchronized states $S\approx0$. For a GAS case (Fig.\ref{epstau}(f)), we find that $x_2(t)$ leads $x_1(t)$ by $\tau_d\approx\rvert\tau_1-\tau_2\lvert$, and also using linear regression between $x_2(t)$ and $x_1(t+\tau_d)$ we find that $x_2(t)=H(x_1(t+\tau_d))\approx0.909x_1(t+\tau_d)$. With this relation, from Eq.\eqref{sim}, we find the similarity function, $S_{GAS}=0.039$; however, if we consider $H$ as an identity function (as in the case of conventional AS) we have  $S=0.102$ that is much larger than $S_{GAS}$, which confirms the occurrence of GAS. At this point it should be noted that, in general, $H$ is not a linear scaling factor (unlike projective synchronization\cite{dghosh}), thus a higher order polynomial regression is needed to describe the form of $H$ more precisely, and that results in a much lower value of $S_{GAS}$. Similarly, for GLS (Fig.\ref{epstau}(h)) we find  $x_2(t)=H(x_1(t-\tau_d))\approx1.081x_1(t-\tau_d)$  ($\tau_d\approx\rvert\tau_1-\tau_2\lvert$) with a similarity function $S_{GLS}=0.035$, which is less than $0.112$, computed by taking $H$ as an identity function (i.e., conventional LS). For $\tau_1=\tau_2$, we observe complete synchronization with $x_2(t)=x_1(t)$, and $S=0$.

\begin{figure*}
\includegraphics[width=.7\textwidth]{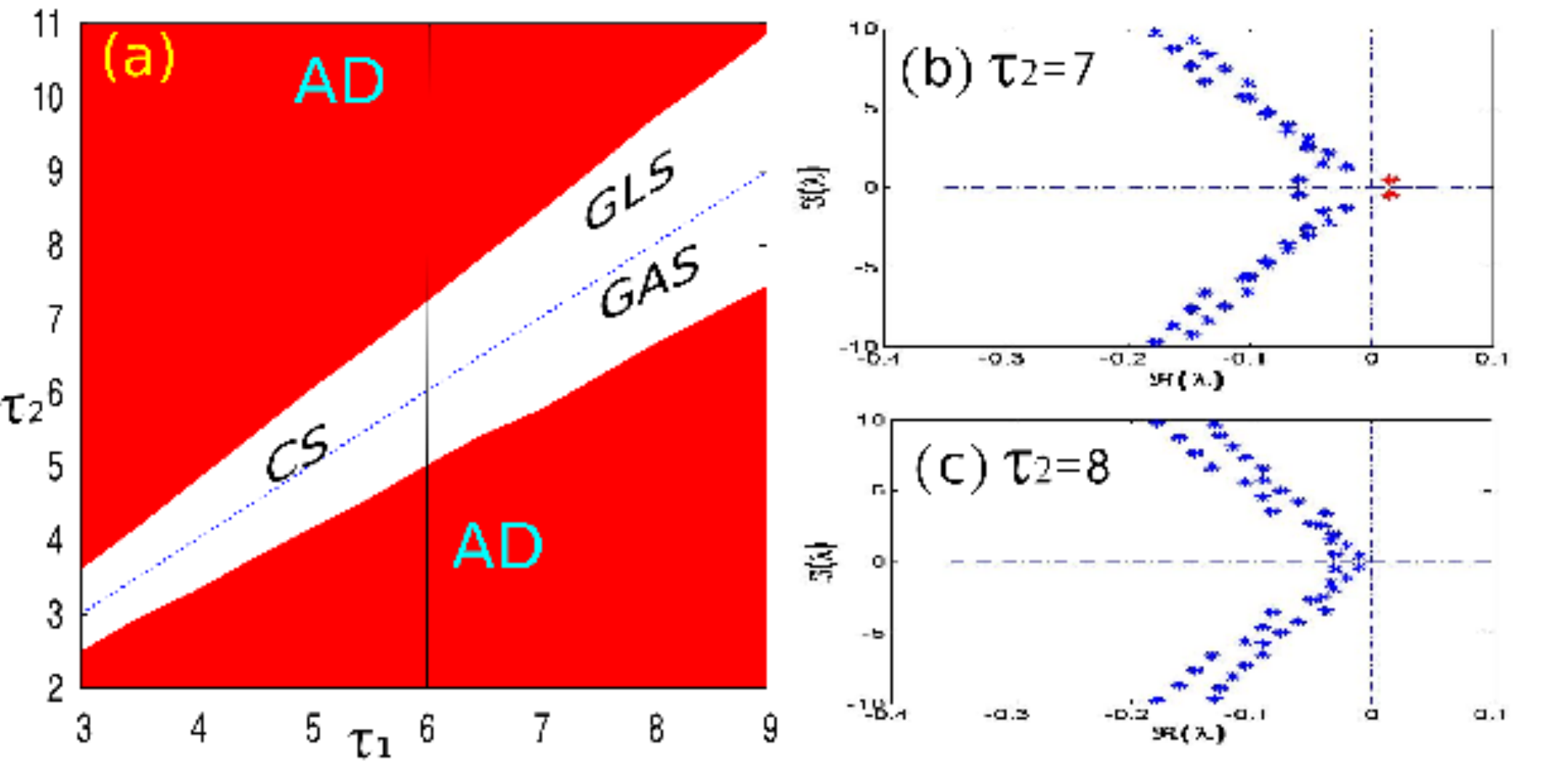}
\caption{\label{tau12} (Color online) {\bf (a)} Phase diagram in $\tau_1$--$\tau_2$ parameter space ($\epsilon=5$, and $Q=0.6$). dotted (blue) diagonal line indicates CS for $\tau_1=\tau_2$. {\bf(b,c)} The eigenvalue spectrum of the coupled systems (b) $\tau_2=7$ (i.e. near but before AD) (c) $\tau_2=8$ (i.e., near but after AD); note that the real part of all the eigenvalues now become negative.}
\end{figure*}

Next, we take a sufficiently high coupling constant ($\epsilon=5$) to ensure synchronized state and vary $\tau_1$ and $\tau_2$ (Fig.\ref{tau12}(a)) simultaneously. We observe that two AD regions are separated by a synchronized state consisting of GAS ($\tau_1>\tau_2$), CS ($\tau_1=\tau_2$, i.e., along the diagonal dotted(blue) line), and GLS ($\tau_1<\tau_2$). Thus, with the variation of $(\tau_1-\tau_2)$ we can clearly observe the transition from AD$\rightarrow$GAS$\rightarrow$CS$\rightarrow$GLS$\rightarrow$AD. We observe that for higher values of intrinsic time-delays, larger mismatch is required to achieve AD for a fixed coupling strength. In this context, we also noticed that, for equal intrinsic time-delays, with increasing intrinsic time-delay, critical value to get AD increases slightly; this fact can be explained from Eq.\eqref{ad}, which shows that for a fixed $Q$, $\epsilon_{cr}$ is proportional to $\delta$ that is a function of intrinsic time-delay.

To confirm the occurrence of AD, we compute the eigenvalue spectrum of the coupled systems using the bifurcation package DDE-BIFTOOL \cite{biftool}. For an illustrative example, Figures~\ref{tau12}(b) and (c) show the eigenvalue spectrum of the coupled systems for $\tau_2=7$ (i.e., near but before AD) and $\tau_2=8$ (i.e., near but after AD), respectively ($\epsilon=5$ and $\tau_1=6$). It can be seen that, with increasing $\tau_2$, the real part  of the largest eigenvalue changes from positive to negative value confirming the occurrence of AD in the coupled system. Further, we observe that  the largest complex conjugate pair of eigenvalues cross the imaginary axis $\Im(\lambda)$ from right to left confirming that the route to AD is through {\it Hopf bifurcation}.

\subsubsection{Effect of coupling: transitions among unsynchronized, PS, CS, and AD}
\begin{figure*}
\includegraphics[width=.65\textwidth]{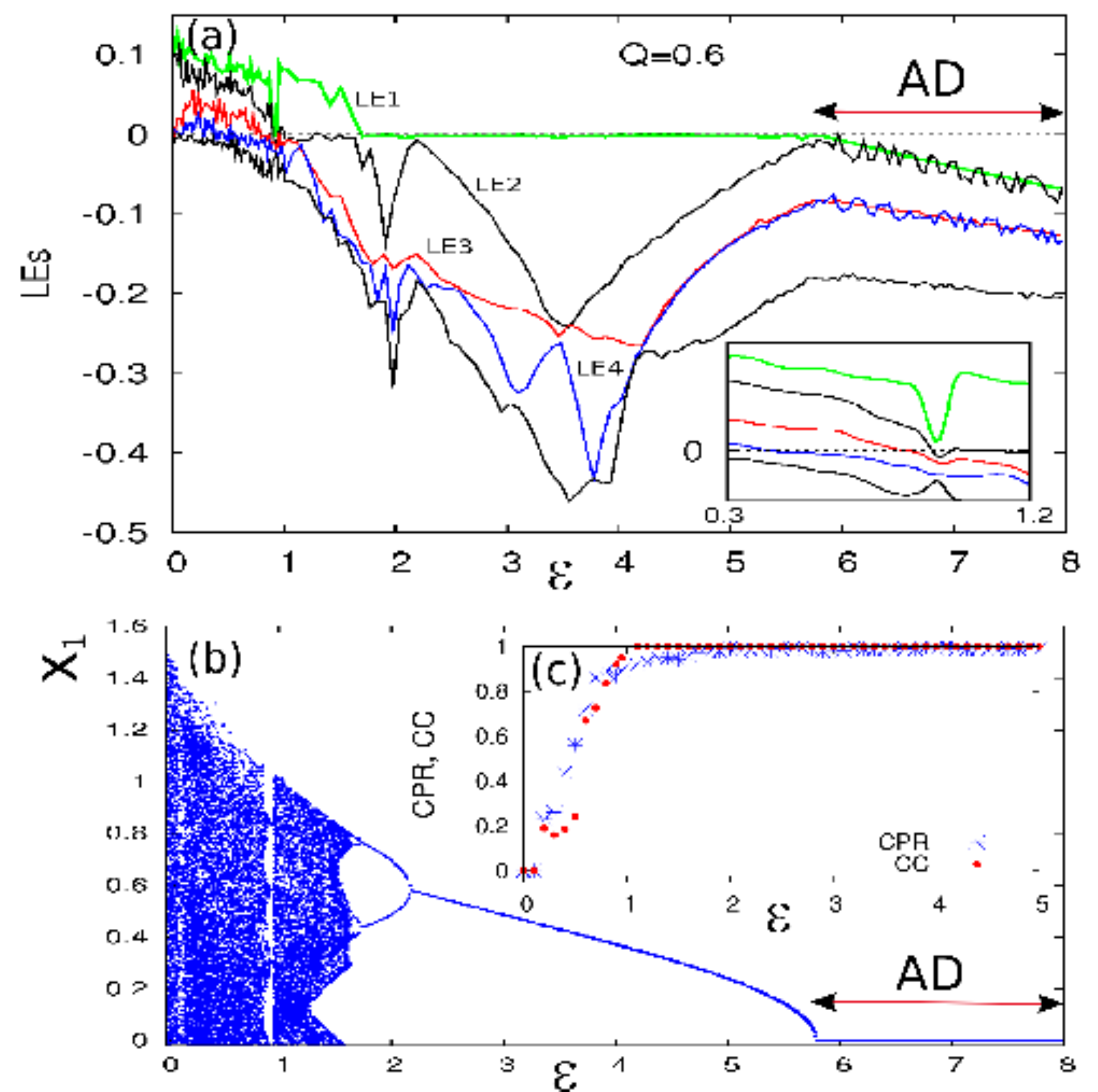}
\caption{\label{lpv_ep} Q=0.6: (a) Lyapunov exponent (LE) spectrum of largest five LEs. Inset shows the same, but now the range is $\epsilon\in(0.3,1.2)$), and curves are smooth (through averaging) for clarity. (b) Bifurcation diagram of $x_1$ with {\ep}. AD represents AD region in {\ep} parameter space. (c) Variation of CPR and CC with {\ep} (Q=0.6).}
\end{figure*}
Next, we set $\tau_1=\tau_2=3$, and vary {\ep} (with $Q=0.6$).  Figure~\ref{lpv_ep}(a) shows the first five LEs in the $\epsilon$ parameter space, $\epsilon \in(0,8)$. Inset of the figure shows the same in $\epsilon \in(0.3,1.2)$, but with smooth curves. It can be seen that LE4 becomes negative at $\epsilon\approx0.65$ that indicates the onset of phase synchronization (PS)\cite{piko}. Further, LE3 makes a transition from a positive to negative value at $\epsilon\approx1$, indicating the onset of complete synchronization (CS). With further increase in $\epsilon$, the largest LE, LE1, becomes zero at $\epsilon\approx1.85$, indicating the fact that the dynamics of the coupled systems now become periodic. The transition of LE1 from zero to a negative value is indicative of AD in the coupled systems. With further increase in $\epsilon$, LE1 monotonically decreases toward a more negative value ensuring the stability of the AD state. The AD state can best be observed from the bifurcation diagram of $x_1$ (Fig.~\ref{lpv_ep}(b)) with $\epsilon$. 

The transition from the unsynchronized state to complete synchronized state through in-phase synchronized state is verified using {\it correlation of probability of recurrence} (CPR), and {\it cross correlation function} (CC). CPR is a quantitative measure of phase synchronization (PS) introduced in Ref. \onlinecite{cpr,*cpr2}. It is related with the  generalized autocorrelation function ($P(t)$), which is defined as,  
\begin{equation}\label{P}
P(t)=\frac{1}{N_1-t}\sum_{i}^{N_1-t}\Theta \left(\epsilon_t-\lVert X_i-X_{i+t}\rVert\right), 
\end{equation}
here $\Theta$ is the Heaviside function, $X_i$ is the $ith$ data point in the $X$ variable, $N_1$ is the total number of data points, $\epsilon_t$ is a preassigned threshold value, and $\lVert . \rVert$ represents the Euclidean norm. CPR is defined as\cite{cpr,*cpr2}: $\mathrm{CPR}=\frac{\langle \bar{P_1}(t)\bar{P_2}(t)\rangle}{\sigma_1\sigma_2}$; ${\bar{P}}_{1,2}$ present that the mean value has been subtracted, and $\sigma_{1,2}$ are the standard deviations of the $P_1(t)$ and $P_2(t)$, respectively. For PS states, $\mathrm{CPR}\approx1$. Further, CC is defined as
\begin{equation}
\mathrm{CC}=\frac{\langle(x_1(t)-\langle x_1(t)\rangle)(x_2(t)-\langle x_2(t)\rangle)\rangle}{\sqrt{\langle\left(x_1(t)-\langle x_1(t)\rangle\right)^2\rangle\langle\left(x_2(t)-\langle x_2(t)\rangle\right)^2\rangle}}.
\end{equation}
CC is a measure of complete synchronization (CS)\cite{piko}; in the CS state, CC$=1$. Figure \ref{lpv_ep} (c) shows the variation of CPR and CC with {\ep}. Increase of both the measures from a zero value with increase in the coupling strength {\ep} agrees with the LE spectrum and bifurcation diagram. For $\epsilon>0.65$, CPR attains values nearly equal to one indicating the onset of PS, and for $\epsilon>1$, CC attains a value of unity indicating the onset of CS in the coupled systems.

\begin{figure}
\includegraphics[width=.49\textwidth]{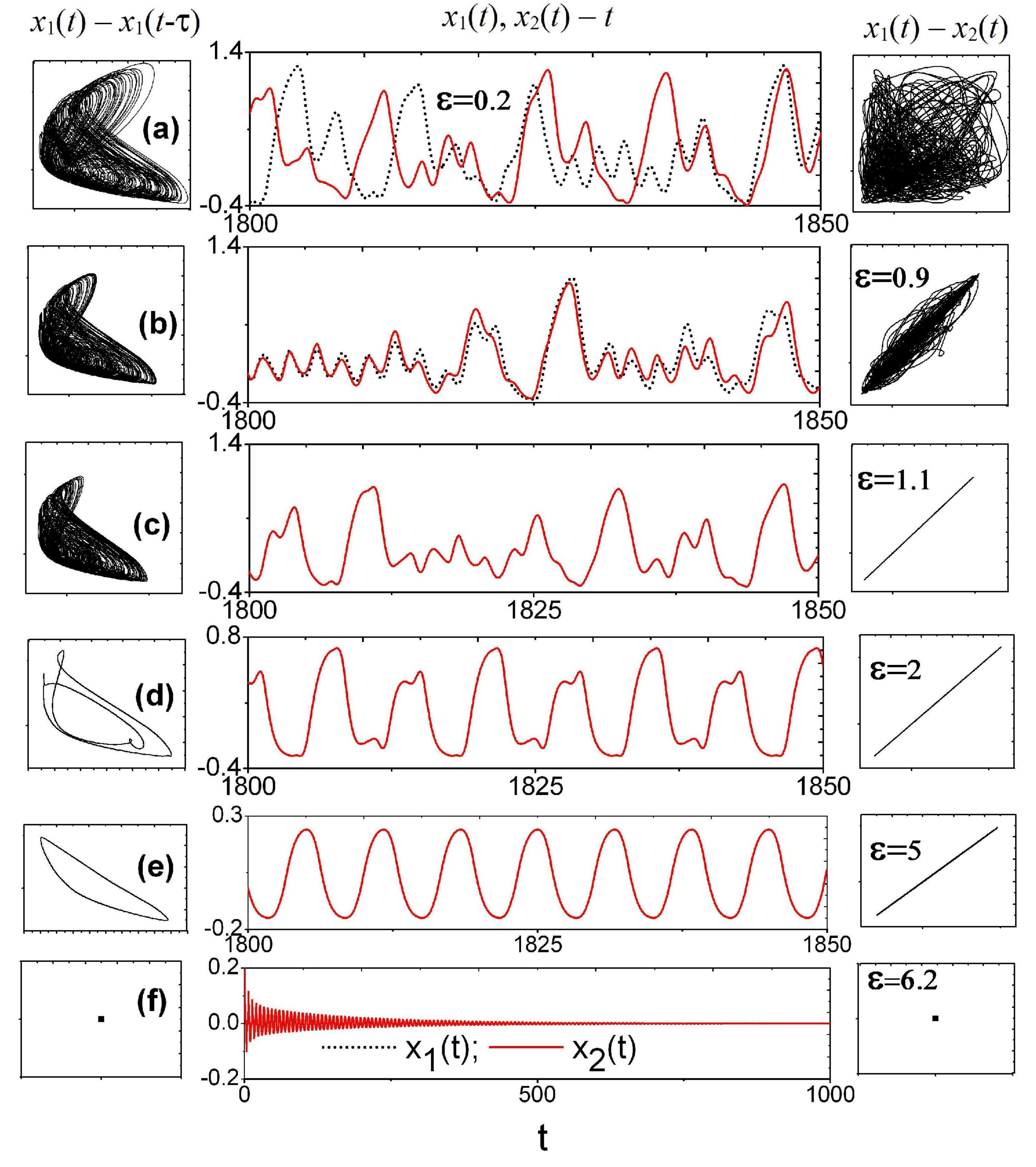}
\caption{\label{rtep} Q=0.6: variation of system dynamics and synchronization states for variable {\ep}. It shows that with increasing {\ep} the coupled systems make a transition from the unsynchronized state (hyperchaotic) (a) to amplitude death state (f) via hyperchaotic in-phase synchronized state (b) and hyperchaotic or chaotic and periodic complete synchronized states (c,d,e).}
\end{figure}

Fig.\ref{rtep} depicts the time variation of $x_1(t)$ and $x_2(t)$ (for $Q=0.6$); it shows that, with  increasing {\ep} the coupled systems make a transition from unsynchronized to complete synchronized states via in-phase synchronized states. Further, this transition is associated with a parallel transition of system dynamics, namely the transition from hyperchaotic to periodic states. AD is shown in Fig.\ref{rtep}(f) for $\epsilon=6.2$, which shows that both of the coupled systems attain the {\it zero} steady state which is the only and trivial steady state of the uncoupled systems. 

\begin{figure}
\includegraphics[width=.49\textwidth]{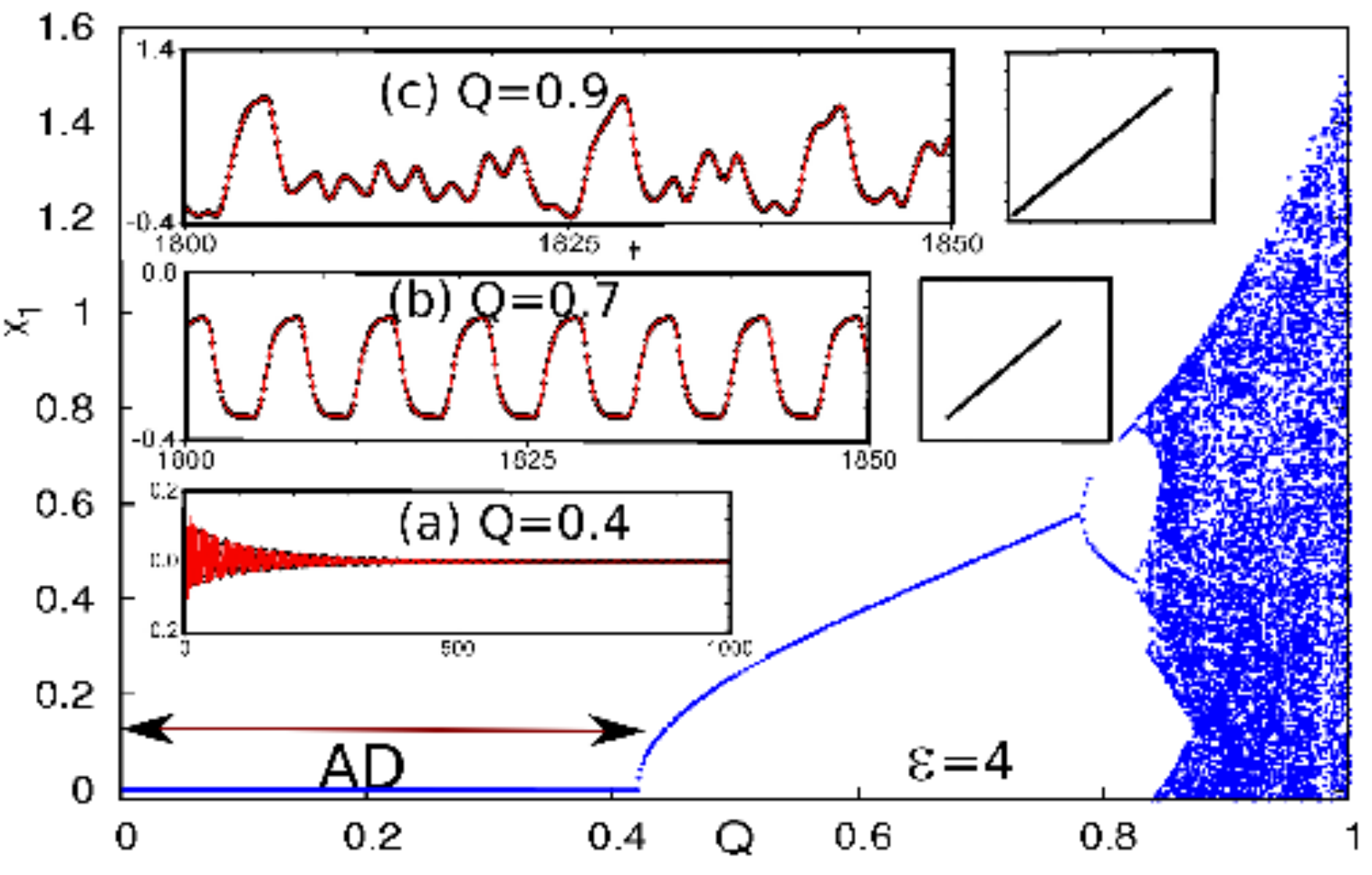}
\caption{\label{varQ} {\ep}=4: Bifurcation diagram of $x_1$ with $Q$. Insets show the time evolution of $x_1$ and $x_2$ (left panel), and $x_1-x_2$ (right panel). (a) AD for $Q=0.4$ (b) Periodic for $Q=0.7$ (c) Hyperchaotic for $Q=0.9$.}
\end{figure}

Next, we keep {\ep} constant at $\epsilon=4$, and vary $Q$, $Q\in(0,1)$. The bifurcation diagram of $x_1$ with $Q$ (Fig.\ref{varQ}) shows that, with increase in the mean-field parameter $Q$, the coupled systems make a transition from the AD state to chaotic and hyperchaotic state through a period doubling route. Figures.\ref{varQ}(a), (b), and (c) show the real time and phase plane plots for three representative values of $Q$ depicting the corresponding transitions among AD, periodic and hyperchaotic states. 

\begin{figure}
\includegraphics[width=.4\textwidth]{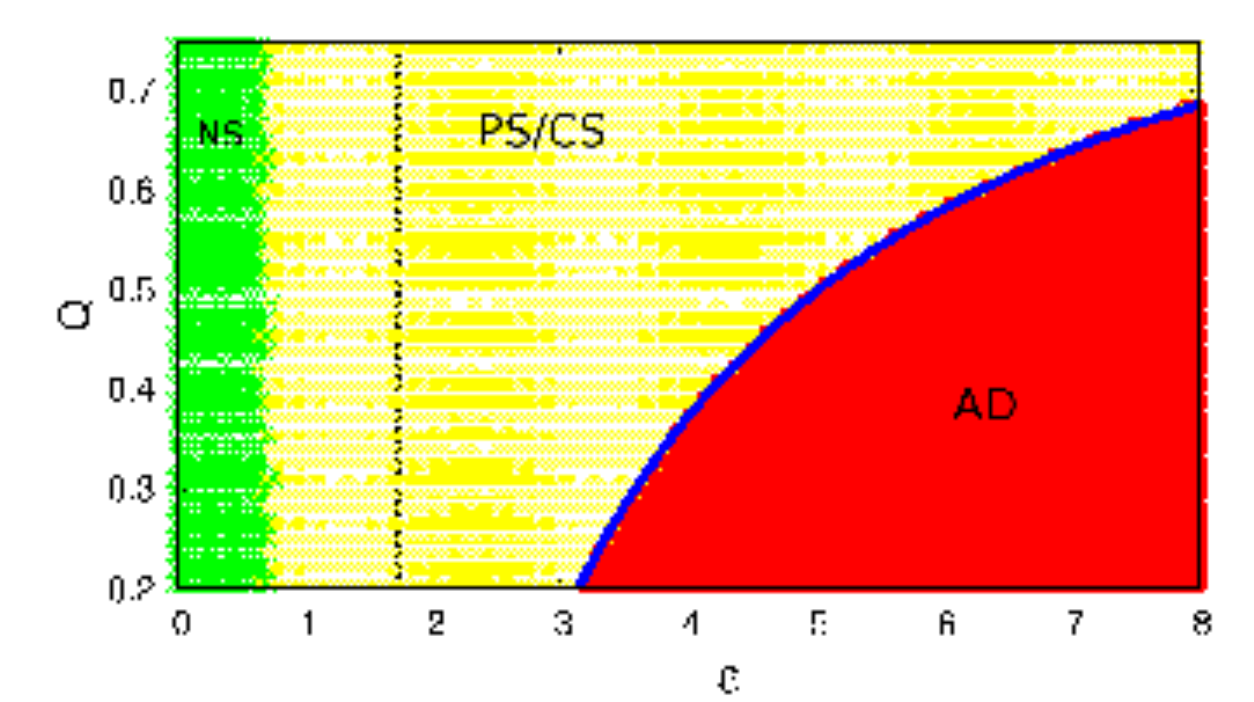}
\caption{\label{phase} (Color online) Phase diagram in {\ep}--$Q$ parameter space. AD: amplitude death; PS/CS: in-phase or complete synchronized state; NS: unsynchronized state. Solid line indicates the analytically obtained critical curve for obtaining AD (with $\delta=2.51$), dotted vertical line indicates the synchronization transition curve.}
\end{figure}

Figure~\ref{phase} depicts the phase diagram in $\epsilon-Q$ space, which shows three distinct regions, namely unsynchronized state (NS), in-phase or complete synchronized state (PS/CS), and amplitude death (AD) state. It is noteworthy that the transition from the unsynchronized state (NS) to synchronized state (PS/CS) does not depend upon the mean-field parameter $Q$, but depends only upon the coupling strength, $\epsilon$, which is in accordance with the analytical result \eqref{cond}. Further, we plot the critical curve (dark solid (blue) line in Fig.~\ref{phase}) for the transition from CS state to AD state using the analytical result \eqref{ad} with an effective choice of $\delta=2.51$; it matches exactly with the numerically obtained critical values in the phase diagram. The effective value of $\delta$ is obtained by fitting numerical results with the analytical result \eqref{ad}. Next, in the phase diagram we show the threshold curve (dotted vertical line) for the transition from NS or PS to CS using analytical result obtained in Eq.\eqref{cond}; since $f^{\prime}(x_{1_\tau})$ (of Eq.\eqref{cond}) is a time varying function, thus, for a general result, here we consider the upper bound of $f^{\prime}(x_{1\tau})$. For $x_{1_\tau}>0$, we have $\lvert f^{\prime}(x_{1_\tau})\rvert_{max}=1.14$; for this value, we show the transition threshold line (dotted vertical line) in the phase diagram. It lies well within the numerically obtained region of CS. For $x_{1_\tau}<0$, we have $\lvert f^{\prime}(x_{1_\tau})\rvert_{max}=2.119$; for this value we have a vertical line (not shown in the figure) at $\epsilon=4.08$ that is also situated well inside the complete synchronized zone in the phase diagram. Since the Krasovskii--Lyapunov theory gives only the sufficient condition of stability of CS thus an exact prediction of the  parameter values for the transition to CS is not possible from Eq.\eqref{cond}. Nevertheless, using this theory we can get a region in the parameter space where CS occurs.

\section{\label{sec:expt}Experiment}
\subsection{\label{sub:elec}Electronic Circuit Implementation}
\begin{figure}
\includegraphics[width=.49\textwidth]{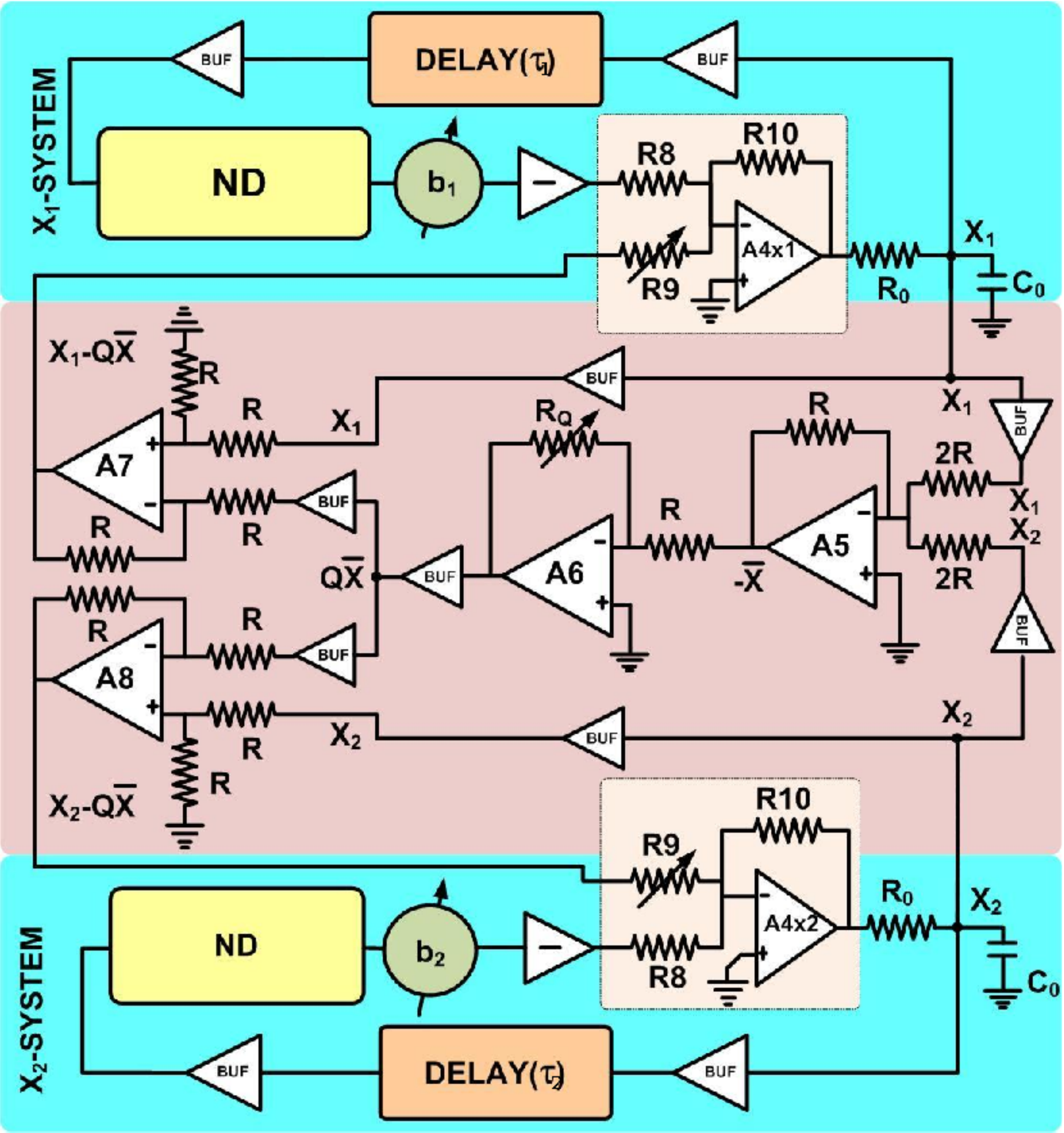}
\caption{\label{mfckt} Representative diagram of the experimental circuit (see text for a detailed description).}
\end{figure}

\begin{figure}
\includegraphics[width=.48\textwidth]{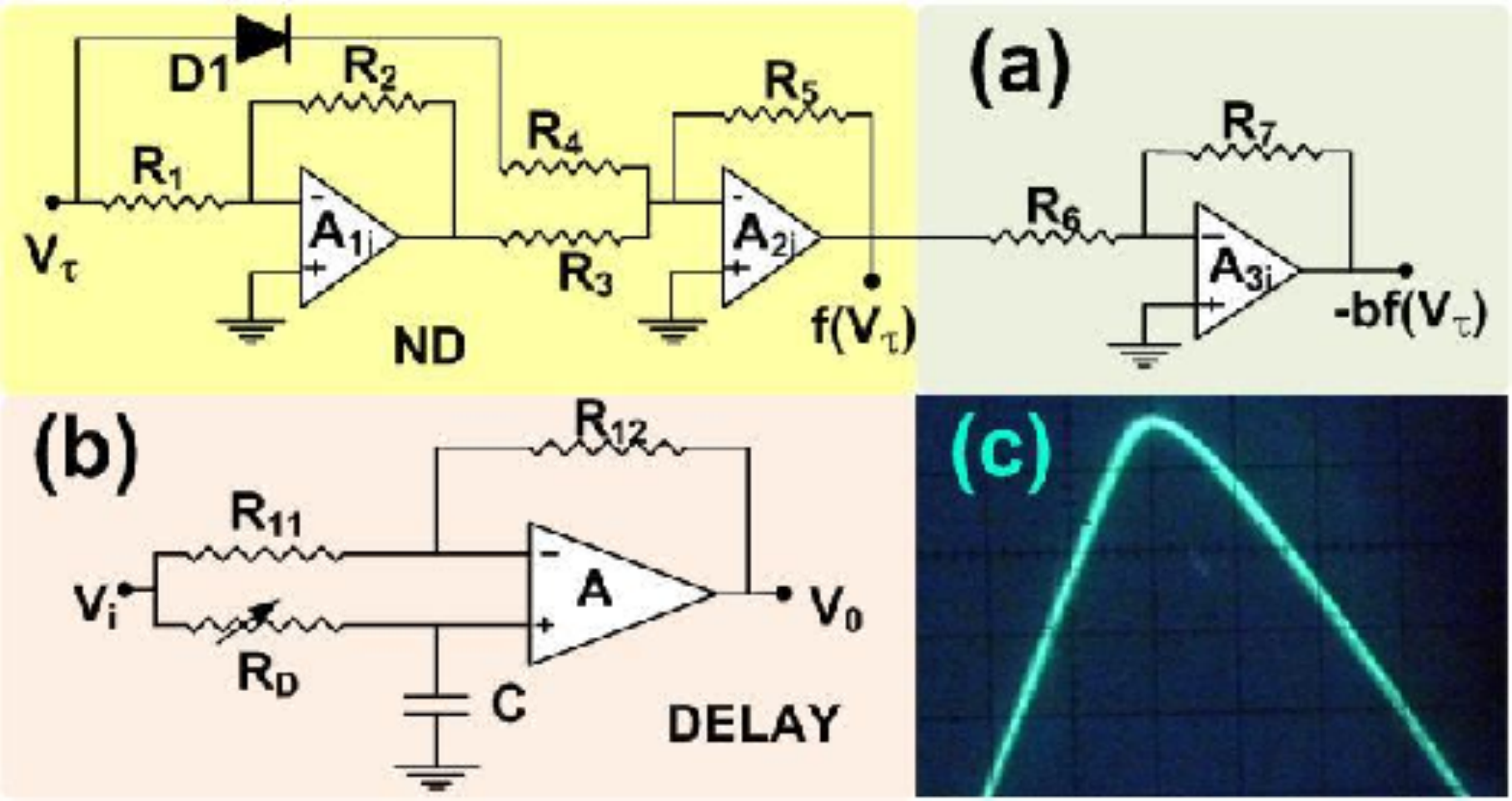}
\caption{\label{nlckt} (a) The nonlinear device (ND) along with $b$. (b) The delay block implemented by active all-pass filter (APF). (c) Experimental nonlinearity of the device. (for the parameter values see text).}
\end{figure}

We set up an electronic circuit level experiment to implement the time-delay system \eqref{syseq} (with \eqref{hw}) under the mean-field diffusive coupling scheme given by Eq.~\eqref{x1x2}. Figure \ref{mfckt} shows the representative diagram of the experimental electronic circuit. The proposed circuit consists of three distinct parts, namely, the $x_1$-system (upper portion), $x_2$-system (lower portion), and the circuit to realize the mean-field coupling (middle portion). Both the $x_1$ and $x_2$-systems consist of a low-pass section ($R_0-C_0$), nonlinear device (ND), delay block (DELAY), gain ($b_1$ and $b_2$), and other circuitry used to realize the proper coupling. The ND block produces the nonlinearity of both $x_1$ and $x_2$-systems; the circuit to realize the ND block is shown in Fig.~\ref{nlckt}(a). For a given input voltage $V_\tau$ (say), this circuit has a nonlinearity, $f(V_\tau)$, given by \cite{banerjee12} 
\begin{equation}\label{nlfn}
\begin{split}
f(V_\tau)&=-0.5\frac{R_5}{R_4}\Big(\lvert V_\tau\rvert+V_\tau\Big)\\
          &~~~~~~+\frac{R_5}{R_3}\beta V_{sat}\tanh{\bigg(\omega\frac{R_2}{R_1}\frac{V_\tau}{V_{sat}}\bigg)}.
\end{split}
\end{equation} 
Here $\beta$ and $\omega$ are certain scaling factors that depend upon the non ideal and asymmetric nature of the op-amps, and $V_{sat}$ is the saturation voltage of the op-amps. The gain part $b_1$ and $b_2$ ($=b$) is realized with op-amp A3 as shown in the same figure. The delay part is implemented using a chain of cascaded active all-pass filters (APF)\cite{Sed03} (shown in Fig.\ref{nlckt}(b)); owing to the almost linear phase response, each delay block produces a time-delay of $T_D\approx R_D C$\cite{banerjee12}. 
Let $V_1(t)$ be the voltage drop across the capacitance $C_0$ of the low-pass section of $x_1$-system, and that of $x_2$-system be $V_2(t)$. Then the equations that represent the circuit dynamics are
\begin{subequations}\label{ckteq}
\begin{eqnarray}
R_0C_0\frac{dV_1(t)}{dt}&=&-V_1(t)-\frac{R_{10}}{R_8}\bigg(\frac{R_7}{R_6}f\big(V_{1_{T_D}}\big)\bigg)\nonumber\\
                        &&~+\frac{R_{10}}{R_9}\bigg(\frac{R_Q}{R}\overline{V(t)}-V_1(t)\bigg),\label{ckteqa}\\
R_0C_0\frac{dV_2(t)}{dt}&=&-V_2(t)-\frac{R_{10}}{R_8}\bigg(\frac{R_7}{R_6}f\big(V_{2_{T_D}}\big)\bigg)\nonumber\\
                        &&~~~+\frac{R_{10}}{R_9}\bigg(\frac{R_Q}{R}\overline{V(t)}-V_2(t)\bigg).\label{ckteqb}
\end{eqnarray}
\end{subequations}
here $f(V_{i_{T_D}})\equiv f(V_{i}(t-T_D))$ ($i=1,2$) is given by Eq.\ref{nlfn}, and $\overline{V(t)}=\frac{V_1(t)+V_2(t)}{2}$.

Now we define the following dimensionless parameters and variables: $t=\frac{t}{R_0C_0}$, $\tau=\frac{T_D}{R_0C_0}$, $x_1=\frac{V_1(t)}{V_{sat}}$, $x_{1_\tau}=\frac{V_{1_{T_D}}}{V_{sat}}$, $x_2=\frac{V_2(t)}{V_{sat}}$, $x_{2_\tau}=\frac{V_{2_{T_D}}}{V_{sat}}$, $\overline{X}=\frac{\overline{V(t)}}{V_{sat}}$, $n_1=\frac{R_5}{R_4}$, $m_1=\beta\frac{R_5}{R_3}$, $l_1=\omega\frac{R_2}{R_1}$, $b=\frac{R_7}{R_6}$, $\gamma=\frac{R_{10}}{R_8}=1$, $\epsilon=\frac{R_{10}}{R_9}$, and $Q=\frac{R_Q}{R}$. With these, \eqref{ckteq} reduces to the following dimensionless form:
\begin{subequations}\label{dleq}
\begin{eqnarray}
\frac{dx_1}{dt}&=&-x_1(t)-bf\big(x_{1_\tau}\big)+\epsilon(Q\overline{X}-x_1),\label{dleqa}\\
\frac{dx_2}{dt}&=&-x_2(t)-bf\big(x_{2_\tau}\big)+\epsilon(Q\overline{X}-x_2),\label{dleqb}
\end{eqnarray}
\end{subequations}
with
\begin{equation}
f(v_\tau)\equiv -0.5n_1(\lvert v_\tau\rvert +v_\tau)+m_1\tanh{(l_1v_\tau)},\label{dlfn}
\end{equation}
where, $v\equiv x_1, x_2$.
Thus, Eq.\eqref{dleq} (with \eqref{dlfn}) is equivalent to Eq.\eqref{x1x2} (with \ref{syseq} and \eqref{hw}) with $a=1$ and proper choice of $n_1$, $m_1$, and $l_1$.
\subsection{\label{sub:exres}Experimental results}
In the experiment the following component values are used: $R_1=10$ k{\ohm}, $R_2=18.55$ k{\ohm}, $R_3=18.55$ k{\ohm}, $R_4=5.6$ k{\ohm}, $R_5=10$ k{\ohm}, $R_6=1$ k{\ohm}, $R_8=R_{10}=1$ k{\ohm}. In the coupling part of Fig.\ref{mfckt}, $R=10$ k{\ohm}. The low-pass sections have $R_0=1$ k{\ohm} k{\ohm}  and $C_0=0.1$ $\mu$F. The APF section of Fig.\ref{nlckt}(b) has $R_{11}=R_{12}=2.2$ k{\ohm}, $C=10$ nF, and $R_D=10$  k{\ohm}. All the op-amps are TL 074 IC (quad JFET op-amp) with $\pm15$ volt power supply. The resistors and capacitors have $5\%$ tolerance. $R_9$ and $R_Q$ are varied with precession potentiometers (POT). With these values, the experimental nonlinearity is shown in Fig.\ref{nlckt}(c), which is same for both the systems. To drive the systems into hyperchaotic zone we use $\tau_{1,2}\ge3$, and $b_1$=$b_2$=$2.4$ by setting $R_7=2.4$ k{\ohm}.

\begin{figure}
\includegraphics[width=.49\textwidth]{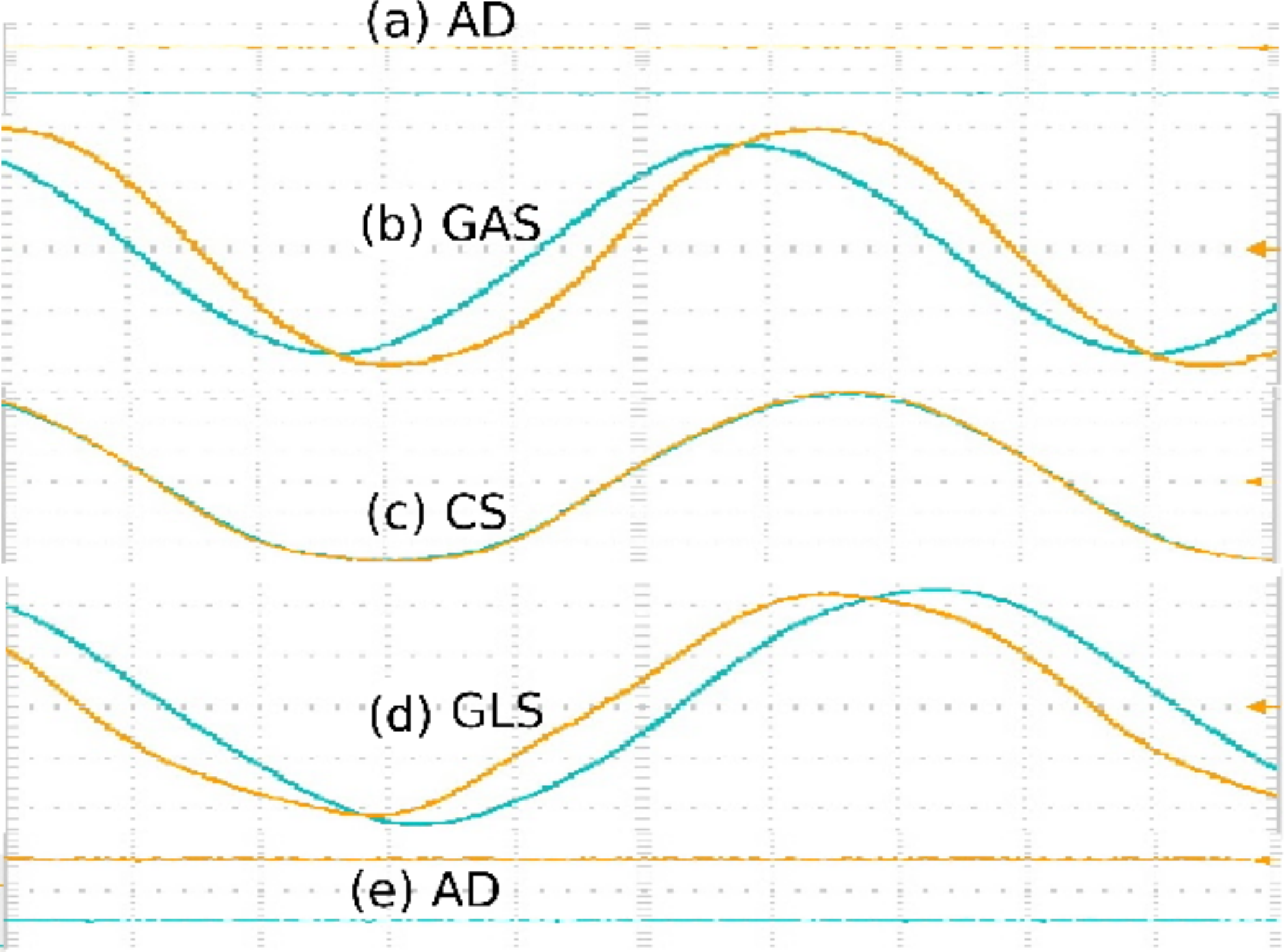}
\caption{\label{vardel} (Color online) Experimental demonstration of transitions among: (a) AD ($\tau_2=4$), (b) GAS ($\tau_2=5$), (c) CS ($\tau_2=6$), (d) GLS ($\tau_2=7$), (e) AD ($\tau_2=8$). $R_9=139$ {\ohm}, $R_Q=8.76$ k{\ohm}, and $\tau_1=6$. $x_1$-system (light gray(orange) trace), $x_2$-system (dark gray(blue) trace). (Scale div: $x$-axis: 12 $\mu$sec/div, $y$-axis:1.25 volt/div).}
\end{figure}

{\it (i) Effect of intrinsic time-delay:} To demonstrate the effect of variation of intrinsic time-delay for a fixed coupling strength, we set $R_9=139$ {\ohm}, $R_Q=8.76$ k{\ohm} and $\tau_1=6$, and vary $\tau_2$. Figure\ref{vardel} shows the transition from AD (Fig.\ref{vardel}a) to GAS ($\tau_2=5$) (Fig.\ref{vardel}b) to CS ($\tau_2=6$) (Fig.\ref{vardel}c) to GLS ($\tau_2=7$) (Fig.\ref{vardel}d), and again to AD (Fig.\ref{vardel}e). It can be seen from Fig.\ref{vardel}(b) that the $x_2$-system (dark gray(blue) trace) leads the $x_1$-system (light gray(orange) trace), and at the same time the waveform of $x_2$ differs from that of $x_1$, both indicate the occurrence of GAS. Fig.\ref{vardel}(d) shows the case of GLS; here  $x_2$ lags behind $x_1$, and waveform of $x_2$ and $x_1$ are different, which is in accordance with the numerical results (Fig.\ref{tau12}). We also observe GAS and GLS in the hyperchaotic zone keeping proper values of $R_9$ and $R_Q$ (not shown here), which indicates that these  phenomena are general.

\begin{figure}
\includegraphics[width=.49\textwidth]{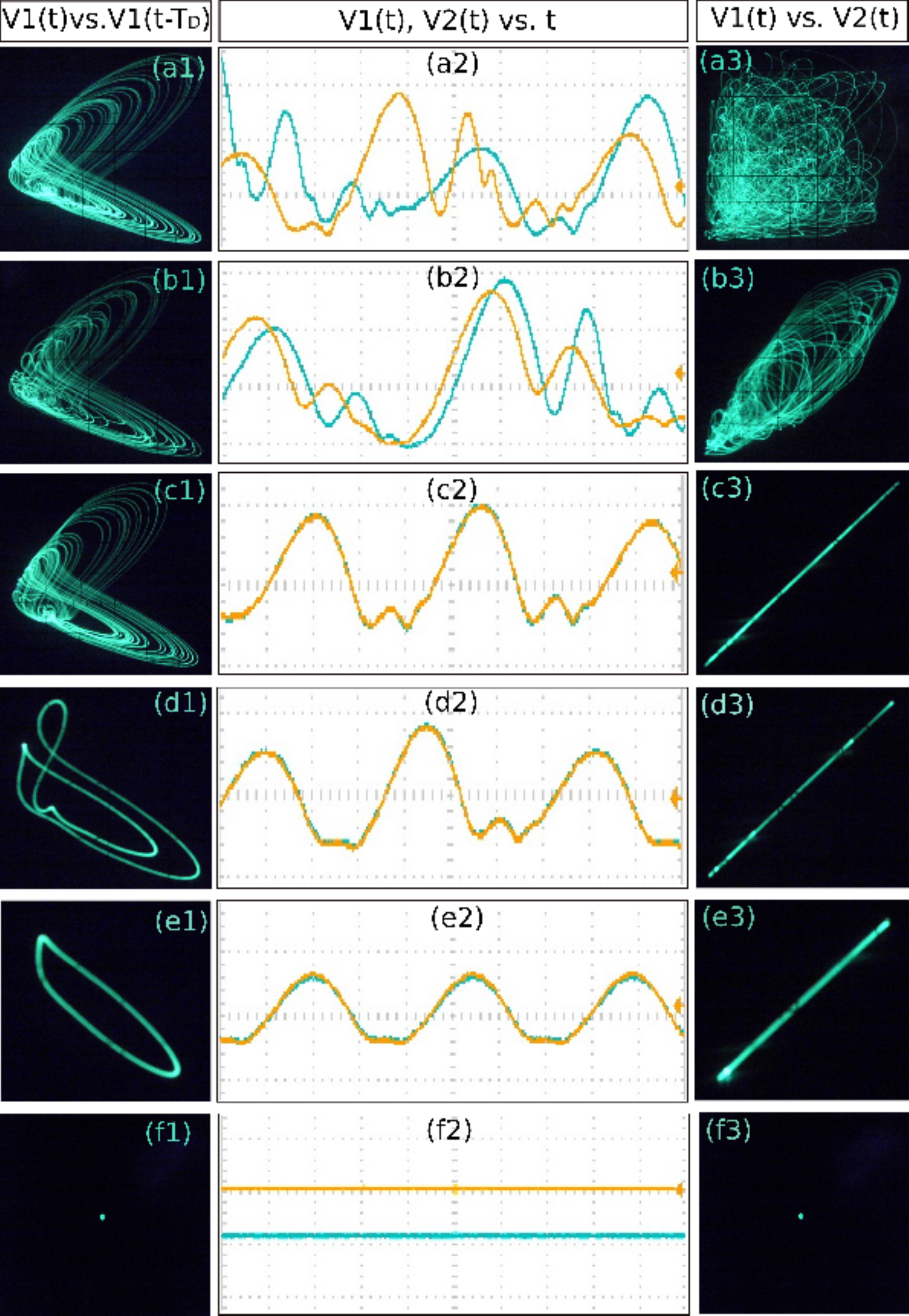}
\caption{\label{vareps}(Color online) Experimental waveforms and phase-plane plots with fixed $Q$ ($R_Q=7.5$ k{\ohm}) and variable $\epsilon$. The left column shows the phase-plane plots ($V_1(t)-V_1(t-T_D)$), the middle column shows the real time waveforms of $V_1(t)$- (yellow) and $V_2(t)$-(blue), and the right column shows the phase-plane plots in ($V_1(t)-V_2(t)$) plane. a(1-3) The unsynchronized state; b(1-3) in-phase synchronization; c(1-3) complete synchronization; d(1-3) period-2 oscillation; e(1-3) period-1 oscillation, and f(1-3) show AD; in (f2) the trace of $V_2(t)$ (blue) is shifted downwards by  1.2 volt from  that of $V_1(t)$ (yellow trace). (For the parameter values see text; Scale div: Second column (a2-f2): $x$-axis: 25 $\mu$sec/div, $y$-axis:1.25 volt/div. Other plots: $x$ and $y$-axes: 0.5 v/div)}
\end{figure}

{\it (ii) Effect of coupling:} We set $\tau_1=\tau_2=3$, $R_Q=7.5$ k{\ohm} and vary $\epsilon$ by varying $R_9$. The results of this variation are shown in Fig. \ref{vareps}. For $R_9=10.77$ k{\ohm}, the scenario is shown in the first row a(1-3) of Fig.\ref{vareps}; (a1) shows the hyperchaotic attractor, and (a2) and (a3) show that there is no correlation between the coupled systems for these parameter values, and both the systems evolve independently. For $R_9=862$ {\ohm}, one can observe in-phase synchronization (Fig. \ref{vareps}(b1-b3)). The third row (c(1-3)) shows the complete synchrony for $R_9=120$. Period-2 (fourth row d(1-3))  and Period-1 oscillations (fifth row e(1-3)) are shown for  $R_9=86$ {\ohm} and $R_9=30$ {\ohm}, respectively. At very low coupling resistance the coupled systems show amplitude death (AD); Fig.\ref{vareps}(f2) shows the waveforms for $R_9=15$ {\ohm} that indicates the occurrence of AD, i.e., now  the oscillations in both the systems die out. 
 
\begin{figure}
\includegraphics[width=.45\textwidth]{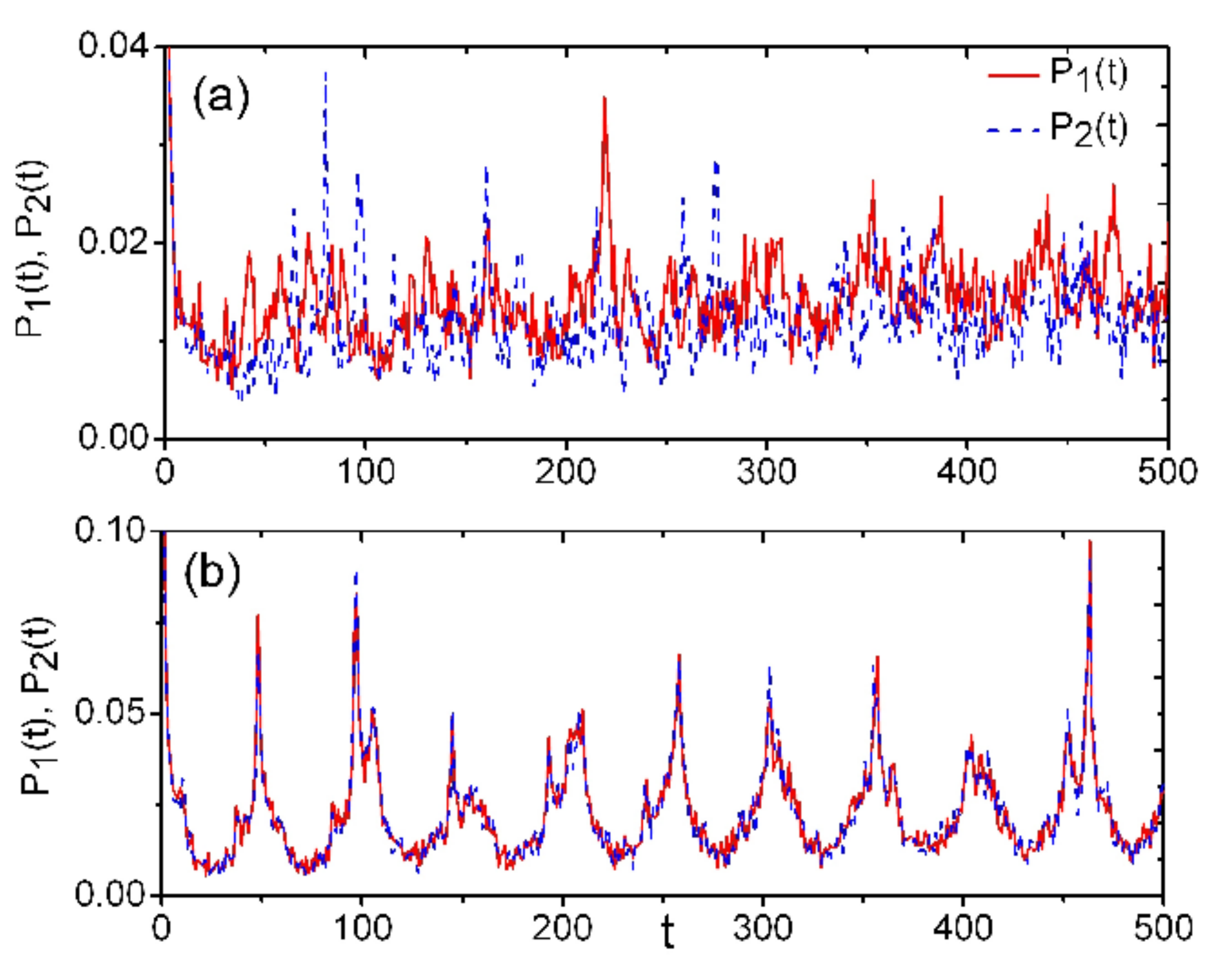}
\caption{\label{cpr_expt} (Color online) Plots of generalized autocorrelation functions using the experimental time-series data (a) unsynchronized state (parameters are the same as Fig.\ref{vareps}(a1-a3)), (b) in-phase synchronized state (parameters are the same as Fig.\ref{vareps}(b1-b3)).}
\end{figure}

$P_1(t)$ and $P_2(t)$ of Eq.\eqref{P} is computed from the experimental time-series data (acquired using DSO, Tektronix TDS2002B, $60$ MHz, 1 GS/s) ($\epsilon_t=0.01$, and $N_1=2400$). Figure~\ref{cpr_expt}(a) shows $P(t)$s for the unsynchronized case, which shows that peaks of $P_1(t)$  does not match with that of $P_2(t)$ in the $t$-axis, indicating unsynchronized states. Figure~\ref{cpr_expt}(b) is for the in-phase synchronization; here the dominant peaks of $P_1(t)$ and $P_2(t)$ matches exactly in the $t$-axis. 
\begin{figure}
\includegraphics[width=.48\textwidth]{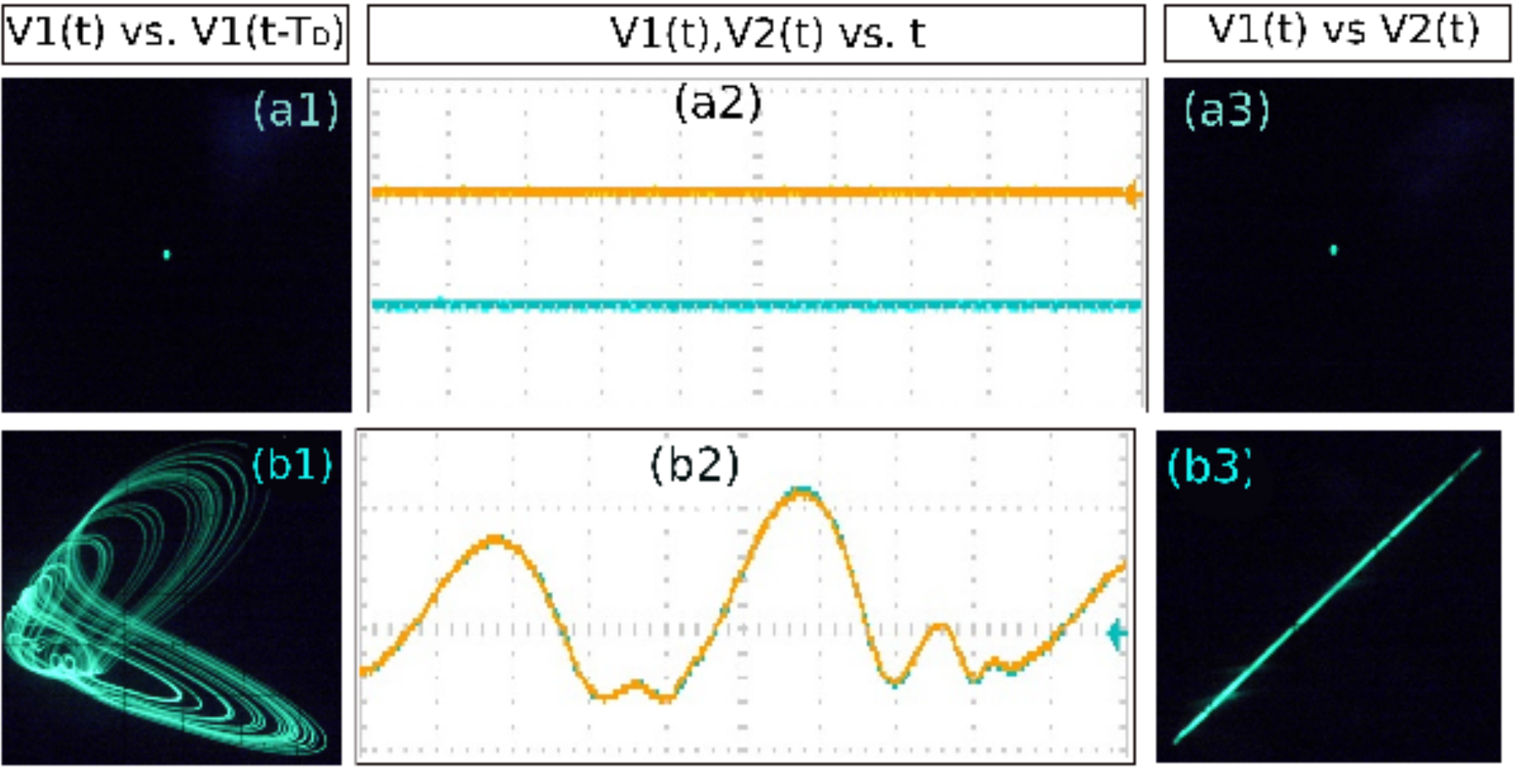}
\caption{\label{varq} (Color online) Experimental plots for fixed $\epsilon$ ($R_9=139$ {\ohm}) and variable $Q$. a(1-3) AD  (the plot of $V_2$ (blue) is shifted downwards by 1.25 volt from that of the $V_1$ for clarity); b(1-3) hyperchaotic oscillation. In all the cases the right column shows that the systems are in CS state. (Scale divisions are the same as Fig.\ref{vareps})} 
\end{figure}

Fig.\ref{varq}a(1-3) show that for a low value of $Q$, AD occurs for $R_Q=8.42$ k{\ohm} ($R_9=139$ {\ohm}). Increase in $Q$ results in oscillation and period doubling scenario. With a large $Q$, both of the systems enter into a chaotic or hyperchaotic zone;  Figs. \ref{varq}(b(1-3)) show this for $R_Q=9.76$ k{\ohm}. These observations are in accordance with the numerical results.

\section{\label{con}Summary and Conclusion}
In this paper we have explored the phenomena of amplitude death and the related synchronization transitions leading to amplitude death in intrinsic time-delayed hyperchaotic oscillators coupled through mean-field diffusion. We have identified two types of synchronization transitions that lead to amplitude death (AD): 

{\it First}, a novel transition scenario, namely  the transitions among AD, generalized (anticipatory, lag) (GAS, GLS) and complete synchronization (CS); {\it this transition is mediated by the variation of the difference of the intrinsic time-delays}, and has no analogue in  coupled low-dimensional systems (with or without coupling delay). 

{\it Second}, transition to  the amplitude death state from an unsynchronized state via in-phase (complete) synchronized states. {\it This transition is mediated by the coupling parameters} (with the coupled systems having equal intrinsic time-delays).

We have derived a stability condition for the GAS, GLS, and CS cases using Krasovskii-Lyapunov theory; also, stability analysis has been carried out to predict the zone of AD in the parameter space. We have exemplified our results numerically using a prototype hyperchaotic oscillator with intrinsic time-delay. Through the modified similarity function, LE spectrum, correlation functions, and eigenvalue spectrum we have identified the zone of GAS, GLS, CS, and amplitude death in the parameter space. It has been found that  numerical results agree well with the analytical derivations. The eigenvalue spectrum of the coupled systems revealed that, in the present system, the route to amplitude death is through Hopf bifurcation. Through the transient dynamics and the Lyapunov exponent spectrum, it has been shown that, unlike systems with {\it coupling time-delay}, the variation of {\it intrinsic time-delay} does not induce {\it phase-flip transition}, but results in transitions among GAS, CS and GLS. Finally, we set an experiment using electronic circuit to demonstrate all the transition scenarios and amplitude death. It has been observed that the experimental results qualitatively agree well with the analytical results and numerical observations. The present study can be extended to the network of mean-field coupled time-delayed systems with distributed intrinsic time-delays, that may reveal the phenomena of GAS and GLS in a more general way.

\begin{acknowledgments}
Authors are grateful to the reviewers for their valuable comments and constructive suggestions; particularly, the suggestion of one of the reviewers to explore the effect of time-delay on the synchronization scenario leads to the present form of the paper. Authors are grateful to Professor B. C. Sarkar for the useful discussions and suggestions. D.B acknowledges the financial support provided by the University of Burdwan, India.
\end{acknowledgments}

\providecommand{\noopsort}[1]{}\providecommand{\singleletter}[1]{#1}%

\end{document}